\documentclass[useAMS,usenatbib]{mn2e}

\usepackage{graphicx}
\usepackage{amssymb}
\usepackage{amsmath}

\title[Glitch detection probability]{On the detection probability of
  neutron star glitches}

\author[M. Yu \& Q.-J. Liu]{M. Yu$^{1,2}$\thanks{E-mail:
    vela.yumeng@gmail.com, meng.yu@nao.cas.cn} and Q.-J. Liu$^3$\\$^1$
  National Astronomical Observatories of China, Chinese Academy of
  Sciences, 20A Datun Road, Chaoyang District, Beijing 100012,
  China\\$^2$ Key Laboratory of Radio Astronomy, Chinese Academy of
  Sciences, 20A Datun Road, Chaoyang District, Beijing 100012,
  China\\$^3$ Department of Astronomy, School of Physics, Peking
  University, Beijing 100871, China}

\begin{document}
\maketitle
\begin{abstract}
Neutron stars are observed to undergo small, abrupt rotational
speed-up. This phenomenon is known as glitch. In pulsar timing
observations, detection of a neutron star glitch is constrained by the
time of occurrence of the event relative to entire observing span and
observing cadences, time of occurrence of preceding/subsequent
glitches relative to observing cadences and the strength of timing
noise. Using the \citet{ymh+13} data sets, in this paper, we analyse
the observational selection in terms of detection probability. We
define partial probabilities for the constraints and use Monte Carlo
method with assuming glitches distribute uniformly to solve the
complete probability formula for both group case involving 157 pulsars
and individual cases for each of the seven pulsars with glitch numbers
$\geqslant 5$. In the simulations, numerical Bayesian inference is
used for glitch identification. With the derived detection probability
density and observed results, we uncover glitch size probability
distribution embedded in the data for both the group and individual
cases. We find the most prominent correction occurred for
PSR~J1341$-$6220, in which exponent of the power-law model varies from
the observed $+0.7^{+1.4}_{-0.7}$ to $-0.4^{+1.0}_{-0.4}$. We suggest
observers determine the detection probability for glitch theories,
e.g. the self-organised criticality.
\end{abstract}
\section{Introduction}\label{sec:intro}
As the rapid co-rotation with the star, neutron star magnetosphere
accelerates charged particles, generates emission and forms radiative
beams. These processes make neutron stars periodic signal emitters. In
an observation of a radio pulsar, received pulsar emission is
integrated to increase detection significance of the pulse so as to
determine its time-of-arrival (ToA). Same (or similar) observation can
be carried out some time later with obtaining another ToA. This
process may be repeated over years such that a series of ToAs is
obtained. Intervals between two ToAs are usually not fixed but vary,
e.g. from minutes, through hours to weeks. The observing cadence is
dependent on artificial observing schedules. In large-scale, long-term
observing programmes, a number of pulsars are thus observed, as those
described by \citet{antt94} and \citet{hlk+04}.

To study neutron star rotation, ToAs measured at the observatory are
converted to the Solar system barycentre (a good enough inertial
reference) and are then converted to rotation phases with an ephemeris
(timing model). If the ephemeris has been refined and is the best,
then the derived phases `connect' with each other\footnote{For an
  idealised ephemeris, the observed phases are integers and the phase
  connection or coherence means the ephemeris derives right phase
  differences. For a practical ephemeris, phase coherence means fit
  for a set of ToAs with the ephemeris converges. In the pulsar
  community, the least-squares method has widely been used for the fit
  and observers have always verified solutions by eye.}; for such a
case, we say we have found the `timing solution'. Thus observations
discretely and unevenly `sample' continuous neutron star rotation. For
each phase, its residual is derived as the difference between the
observation and prediction (integer value) made by the ephemeris. The
operations have computationally been realised with a high precision; a
representative is the software package \textsc{tempo2}
\citep{hem06,ehm06}. As the free conversion between ToAs and rotation
phases, the predicted integer phase is equivalent to idealised pulse
arrival times when timing solution is established and phase residual
is equivalent to timing residual. Researchers
\citep[e.g.][]{antt94,dmh+95,hlk10} have found, for most normal and
some recycled pulsars, even with best available ephemerides, after
modelling a steady stellar slowdown, timing residuals are not white
noises but exhibit various random walk-like behaviours. The behaviours
are known as `timing noise'. In power spectrum, timing noise is
recognised with the feature of low frequency excess or `red' such that
power-law functions have been used in modelling
\citep{chc+11,lah+14}. Researchers
\citep[e.g.][]{ywml10,elsk11,ymh+13} have also found that, since a
particular ToA, phases sharply grow faster than predictions and the
coherence is usually (even severely) broken. This indicates a sudden
increase in neutron star rotational rate, glitch, occurred sometime
within the interval defined by the ToA (post-glitch first ToA) and the
one before it (pre-glitch last ToA). In the literature, almost 500
glitches in hundreds of pulsars have been reported.

Observations for 165 normal pulsars between 1990 and 2011 were
searched for glitches by \citet{ymh+13}. At the Parkes Observatory,
observing sessions were scheduled in cadences every 2--4 weeks. In
each session, most member pulsars were observed for 1--10\,min with a
ToA obtained. Thus data spans for the pulsars ranged between 5.3 and
20.8\,yr. Due to artificial changes of the observing projects, some
member pulsars were not observed for some periods over the decades
leaving data gaps in their ToAs. To study the evolution of neutron
star rotational rate, the authors derived pulsar pulse frequencies
$\nu$ by consecutive local fits to the ToA data; number of ToAs per
fit was typically five or six. Each glitch was verified by recognising
a step in the frequencies $\Delta\nu$. To initially measure the glitch
size, frequency values extrapolated with pre- and post-glitch timing
solutions were compared at a glitch epoch. In many cases, the authors
could not determine glitch epoch by assuming change of pulse phase at
the glitch is zero. Instead they assumed glitch epochs as the
mid-point between the pre-glitch last and post-glitch first ToAs with
fitting for a phase shift at the assumed epoch (to assure the phase
continuity). With refining measurements of glitch size and other
glitch parameters by fitting ToAs around glitch with a glitch model,
the authors reported out of the 165 pulsars 36 were seen to glitch and
a total of 107 glitches were identified.

An interpretation for the glitch phenomenon is the avalanche release
of angular momentum from neutron star inner crust superfluid when
differential rotation between the superfluid and the rest of the star
reaches a critical point and pinning force cannot hold the neutron
superfluid vortices any more \citep{wm08,mpw08}. This view is the
derivative of the phenomenon, the `self-organised criticality'
\citep[][and the references therein]{btw88}, that has widely been seen
in nature. The self-organised criticality refers to dissipative,
non-linear systems, in which ordered spatiotemporal structures develop
spontaneously with sustaining an equilibrium to perturbations. It has
spatiotemporal fingerprints: the spatial scale-invariant, self-similar
(fractal) behaviours and the temporal $1/f$ noise or flicker noise. In
many self-organised systems, time spent in building up critical states
is greater than the time scales of avalanche relaxations. Avalanches
occur on all time scales, flicker noise correlates on all time scales
and its power spectrum presents a power-law behaviour with roughly
minus one the power exponent $b$. The $1/f$ noise means a power-law
distribution with exponent $a$ of the duration (lifetime) of
avalanches. So it is not noise but reflects the physics of extended
dynamical systems. Concurrently avalanches occur on all self-similar
spatial scales. Sizes exhibit power-law distribution with exponent $s$
and scale with lifetimes with positive power exponent $c$. The
exponents $s$, $a$, $b$ and $c$ are related to each other through
`scaling laws'. Correlation of spatial scales is described by a
generalised Kolmogorov spectrum. Turbulence is the special case of the
self-organised criticality where self-similarity occurs in both space
and time.

By studying glitch size distributions of nine frequent glitching
pulsars, \citet{mpw08} found the distributions could be fairly well
modelled by power-law functions with exponents falling between $-2.4$
and $+0.13$. Because glitch sizes imply spatial scales of superfluid
avalanches, they suggested glitch phenomenon is a manifestation of the
self-organised criticality. In addition, \citet{mpw08} have also found
the inter-glitch times of seven (out of the nine) pulsars distributed
exponentially exhibiting the evidence of Poisson processes;
distributions of the other two (the Vela pulsar and PSR~J0537$-$6910)
contained Dirac components reflecting quasi-periodicities.

In principle a measurement of pulse frequency requires two ToAs. If a
glitch occurs within the first or last ToA interval, we can neither
verify it (by examining pulse frequencies) nor meausre it (see Section
\ref{sec:prob} for details). There exists a `detection window'. We
define it as a period which covers the entire observing time span but
the first and last ToA intervals. As observations are often seperated
by weeks, we cannot exclude the possibility that two or more glitches
occur coincidentally between two observations. The third factor that
affects the detectability of glitches is researchers
\citep[e.g.][]{dmh+95,wbl01,js06} argued glitches especially small
ones may `dissolve' into timing noises (though this has not been
quantified). These observational selection may have biased the
observed statistics and our understanding to glitch mechanism. A full
analysis to glitch distribution should contain both observations and
detectabilities. In this work, we use the \citet{ymh+13} data sets to
explore the detectability in terms of detection probability. In the
following section, we describe the probability on detecting glitch
events, writing down complete probability formula. Next, in Section
\ref{sec:routn}, we describe our routines used for solving the
probabilities. We show the results of various examinations. In Section
\ref{sec:solun}, we give our solution to the complete probability
formula in the form of probability density. We discuss the
implications in Section \ref{sec:disn}. Then we close with a
conclusion in Section \ref{sec:conc}.

Before starting, we would like to supplement two points here. First,
our results are only for the \citeauthor{ymh+13} data. For instance,
same glitch may occur in the first ToA interval in our data and is
undetectable but may occur in e.g. the third ToA interval in a data
set with more rapid cadences and is detectable. Furthermore, white
noise level might be considered when studying glitch detectability for
recycled pulsars. Second, in the \citeauthor{ymh+13} data, we have
seen glitches usually exhibit several observables. Apart from the
frequency step, pulse frequency first time derivative $\dot\nu$ also
has steps, often negative sometimes positive. Following glitches,
there sometimes show exponential and/or linear recoveries (steps in
$\ddot\nu$). Despite these, out of two reasons, we have made
simplification by only involving frequency step into our analysis: 1)
The numerical Bayesian inference used is computationally expensive, we
could not expand the dimension of parameter space further or the
experiment could not be accomplished in a reasonable time scale with
our available computing facilities (see Section \ref{sec:solun}); 2)
Again, frequency step is the parameter that implies avalanche spatial
scale. The existence of the other observables affects the minimum
detectable glitch in a given data set \citep[see][for an analysis to
  $\dot\nu$ negative step]{eas+14}, while we have restricted our
discussion within the minimum ($1.65\times10^{-9}$\,Hz) and maximum
($3.52\times10^{-5}$\,Hz) glitches detected in the \citeauthor{ymh+13}
data.

Let us begin.

\section{The probability}\label{sec:prob}
For a given set of ToAs, detectability of a glitch event is
constrained by its occurring time (`epoch term $C_{\rm epoch}$'), if
other glitches have occurred in the same ToA interval (`multi-glitch
term $C_{\rm multi}$') and the level of timing noise (`noise term
$C_{\rm noise}$'). Thus we can write the complete probability for a
detection with size $\Delta\nu$ as
\begin{equation}
{\rm P}\{D(\Delta\nu)\} = \left\{ \begin{array}{ll} {\rm
    P}\{D(\Delta\nu)|C_{\rm epoch}\}{\rm P}\{C_{\rm epoch}\}\,+ &\\
  {\rm P}\{D(\Delta\nu)|C_{\rm multi}\}{\rm P}\{C_{\rm multi}\}\,+ &\\
  {\rm P}\{D(\Delta\nu)|C_{\rm noise}\}{\rm P}\{C_{\rm noise}\}
  &\\ \textrm{\qquad\qquad\qquad within detection window} &\\ 0
  &\\ \textrm{\qquad\qquad\qquad otherwise}\label{eq:prob}
\end{array} \right.
\end{equation}
(where, explicitly, the constituents are assumed to be
independent).

For the \citeauthor{ymh+13} data sets, glitch size was initially
measured by comparing the pulse frequencies extrapolated from pre- and
post-glitch timing solutions at the assumed glitch epoch. Refined
solution was obtained by fitting for the glitch model (refer to
equation (1) in the \citeauthor{ymh+13} paper) to the local data
across the glitch. These fitting was realised by \textsc{tempo2} that
implements linear least-squares method (a realisation of the
`frequentist' method for data modelling). Although in principle two
ToAs determine a pulse frequency measurement, in practice, since the
observed phase samples are relative values, an arbitrary phase is
involved in the fit and at least three ToAs are required in both pre-
and post-glitch data span when fitting for the frequency (such that
the initial evaluation for glitch size can be carried out). Therefore,
we shrink the detection window by one ToA interval from both the start
and the end of the entire observing time span. In other words, we
re-define the detection window as the period that covers the entire
observing time span but the first and last {\it two} ToA intervals. We
also define
\begin{equation}
{\rm P}\{C_{\rm epoch}\} = 1 - \frac{{\rm time\,span\,of\,the\,
    first\,and\,last\,two\,ToA\,intervals}}{{\rm
    total\,observing\,time\,span}}\label{eq:p_e}
\end{equation}
as the probability for a glitch epoch to locate in the detection
window with defining
\begin{equation}
{\rm P}\{D(\Delta\nu)|C_{\rm epoch}\} = 1.\label{eq:p_d|e}
\end{equation}
Now let us assume a glitch has occurred in some ToA interval (within
detection window certainly) that spans $\Delta T$. Here comes another
glitch, the probability for it to occur in the ToA interval where the
first glitch is is $\frac{\Delta T}{T}$ and so
\begin{equation}
{\rm P}\{C_{\rm multi}\} = \left\{ \begin{array}{ll}
\frac{\Delta T}{T} & \textrm{for coincidence}\\
1 - \frac{\Delta T}{T} & \textrm{otherwise,}\label{eq:p_m}
\end{array} \right.
\end{equation}
where $T$ is the time span of detection window. Then we define
\begin{equation}
{\rm P}\{D(\Delta\nu)|C_{\rm multi}\} = \left\{ \begin{array}{ll}
0.5 & \textrm{if coincidence}\\
1 & \textrm{otherwise,}\label{eq:p_d|m}
\end{array} \right.
\end{equation}
though actual measurement would be dominated by the larger one if, for
example, the two glitches have sizes $10^{-5}$ and $10^{-7}$\,Hz
respectively. In practice, two glitches can be resolved only if they
are separated by at least three ToAs. Hence it would be better for us
to expand the $\Delta T$ by further including two ToA intervals on
each side about the first glitch. For the \citeauthor{ymh+13} data
sets, $\frac{\Delta T}{T}$ is typically $\sim\frac{120\,{\rm
    d}}{4000\,{\rm d}}$, which is three per cent. Then another three
per cent will be multiplied to give the probability for another glitch
to occur in the ToA interval where the two glitches are. This is
negligible. So, for the multi-glitch term, we only consider the dual
glitch case.

Finally, let us look at the noise term. The piece ${\rm P}\{C_{\rm
  noise}\}$ indicates the level of (timing) noise. If (merely) a
particular pulsar is studied, this piece will then be an arbitrary
value and will become vanished in a normalisation process. In this
case, ${\rm P}\{D(\Delta\nu)|C_{\rm noise}\}$ forms an array (one
dimensional) in which a certain member (bin) indicates the probability
of detections with sizes binned to the specific size interval. For the
case when a bunch of pulsars is studied, e.g. we are studying
\citeauthor{ymh+13} pulsars, the ${\rm P}\{C_{\rm noise}\}$ piece
turns to imply a distribution of noise level and, ${\rm
  P}\{D(\Delta\nu)|C_{\rm noise}\}$ forms a matrix (two dimensional)
in which an element (cell) indicates the probability of detections
distributed into the specific size interval and noise level
interval. Unlike the epoch and multi-glitch terms, the noise term
cannot be determined analytically. We have run a Monte Carlo
simulation to determine it. For either the individual or group case,
the member detection probability is defined as the ratio of the
number of detected events to the total event number distributed into
the bin or cell. Results will be presented in Section
\ref{sec:solun}. Next, we shall describe our numerical routines for
simulating and modelling data sets, which have supported the
simulation.

\section{Routines}\label{sec:routn}
\subsection{Simulating and modelling data}\label{sec:routn:data}
Here, we would like to give an example, by which one can get all ideas
on how have we simulated and modelled data sets. Now let us imagine we
are studying a pulsar, PSR~J0908$-$4913, a member in the
\citeauthor{ymh+13} list. We are going to simulate a set of timing
residuals for it and then model the data with deriving model
parameters.

We start with finding timing solution for its real ToAs. We use
\textsc{tempo2} and fit pulse frequency $\nu$ and the first time
derivative $\dot\nu$ to the data to form phase-connected timing
residuals. We also fit the second derivative $\ddot\nu$ since the
residuals further show an evident cubic structure. As we obtain
phase-connected timing residuals, the pulsar ephemeris is
refined. This is actually what \citeauthor{ymh+13} have done. Then,
idealised pulse arrival times are derived by simply subtracting the
arrival time residuals from the ToAs. We next convert the idealised
arrival times into (relative) integer pulse phases using the pulsar
ephemeris. Results are recorded. With the refined ephemeris and
integer phases, we now begin to produce a simulated data set. This
means we superimpose a designed residual onto each of the integer
phases. As the integer phases are held fixed, even a faulty timing
model can derive exact arrival time residuals and thus no phase
incoherence occurs in any design. Under this principle, we firstly
generate a raw time series by simply giving each integer phase a
Gaussian distributed random number; uncertainty for each of the
simulated ToAs takes that of the real ToA. Ingredients may be added
then. One of them, an important feature for normal pulsars, is the
timing noise. As described by \citet{chc+11}, a timing noise sequence
and the spectrum with amplitude
\begin{equation}
n\sqrt{\frac{P(f)}{T}} = n\sqrt{\frac{A}{T}}[1+(\frac{f}{f_{\rm
      c}})^2]^{-\frac{\alpha}{4}}\label{eq:pden}
\end{equation}
form a Fourier transform pair, the power-law function $P(f)$ is the
sequence's power spectral density, which describes the `red'
feature. In the equation, $n$ is the ToA number, $T$ here indicates
the entire time span of the data set, $f_{\rm c}$ and $\alpha$ are
spectral corner frequency and exponent respectively, and $A$ the
amplitude of the spectral density at $f=0$ characterising the strength
of red noise. To generate red noise, we sample the (amplitude)
spectrum evenly from zero frequency to Nyquist frequency
($\frac{n}{2T}$) with a step size $\frac{1}{100T}$ (one hundred times
finer than the discrete Fourier transform step $\frac{1}{T}$). Then
the real part takes the product of the sample value and a standard
Gaussian distributed random number, so does the imaginary part. We
subsequently do complex-to-real Fourier transform followed by the
Catmull-Rom interpolation to obtain the time series with red noise
feature. To add a frequency jump $\Delta\nu$ (glitch), we simply shift
integer phases after a glitch epoch $t_{\rm g}$ by
$-\Delta\nu(t-t_{\rm g})$. For the constant phase jump at $t_{\rm g}$,
we freeze it at zero. Up until here, the simulation is done.

Next, we shall model the simulated data with deriving model
parameters. We view this problem as a `Bayesian'. To find out the best
agreement between data and model, Bayesians assess the plausibility of
hypotheses (models), rather than merely deriving likelihood of data
for an assumed model as frequentists do. A recent improvement for
computational Bayesian inference has been made by \citet{fhb09} as
\textsc{multinest}, and an interface for pulsar timing analyses has
been developed by \citet{lah+14} as \textsc{temponest}. We use them in
this work. The problem we are studying requires to derive the
posterior probability distributions for each of the parameters in the
parameter space. Parameter estimates are then drawn from the posterior
distributions using standard Markov Chain Monte Carlo
method. According to Bayes' theorem, the posterior probability
distribution is calculated as the ratio of the product of the Bayesian
likelihood and the prior probability distribution to the Bayesian
evidence, a normalising factor presented as the average of the
likelihood over the prior. For the prior distribution of every
parameter in the parameter space, \textsc{multinest} adopts uniform
distribution and we set the boundaries to well cover plausible
values. We extend the parameter space with glitch epoch, glitch size,
the red noise parameters (defined in equation \eqref{eq:pden}) and the
white noise parameters, EFAC and EQUAD\footnote{These parameters are
  designed for artificially changing ToA uncertainty to study two
  white noise components. The EFAC accounts for man-made sources,
  e.g. the various radiometer noise level of observing systems. The
  EQUAD accounts for pulsar intrinsic sources, e.g. the `jitter'
  phenomenon \citep{lkl+12}. Please refer to equation (10) in
  \citet{lah+14} for ToA uncertainty with the EFAC and EQUAD adopted
  in \textsc{temponest}.}, for each flagged observing system. Since we
set the phase jump at glitch epoch as zero, this term is not included
into the parameter space. We marginalise all pulse parameters namely
$\nu$, $\dot\nu$ and $\ddot\nu$. For one sampling process in the
chain, \textsc{multinest} takes a sample in the parameter space,
evaluates the likelihood with the data and calculates the evidence. In
likelihood evaluation, glitch is subtracted by shifting pulse phases
after the sampled glitch epoch $t_{\rm g,s}$ by $\Delta\nu_{\rm
  s}(t-t_{\rm g,s})$ where $\Delta\nu_{\rm s}$ is the sampled glitch
size. The glitch search is realised in this way. Evidence calculation
is computational expensive.  \citet{fhb09} developed the ellipsoidal
nested sampling method to improve the efficiency as well as the
robustness, for the details, please refer to the reference. For the
details on pulsar timing likelihood, please refer to \citet{lah+14}.
\subsection{Examinations}\label{sec:routn:exam}
\begin{figure}
\begin{center}
\includegraphics[angle=-90,width=8cm]{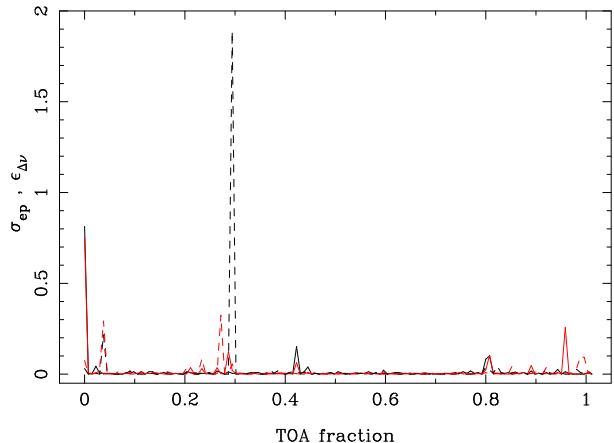}
\end{center}
\caption{Variations of $\sigma_{\rm ep}$ and $\epsilon_{\Delta\nu}$ as
  a function of the fraction of trial glitch epoch to the span of
  detection window. Black lines are for $\sigma_{\rm ep}$, red lines
  are for $\epsilon_{\Delta\nu}$. Solid lines indicate test three,
  dashed lines indicate test four (see text).}\label{fig:edge}
\end{figure}
After describing principles of our routines, let us continue our
example with some actual numbers. We would like to see if
\textsc{temponest} may accurately return glitch epoch and size if the
event is evident to eye, and if the response is uniform over the
detection window. In fact, this is a point we need to examine before
implementing the Monte Carlo simulation (to determine ${\rm
  P}\{D(\Delta\nu)|C_{\rm noise}\}$). We did four tests. In test one,
we simulated sets of timing residuals for PSR~J0908$-$4913 with,
3150\,d time span, 30\,d (evenly) spaced ToAs and 1\,$\mu$s (fixed)
ToA uncertainty. A glitch with $\Delta\nu=10^{-7}$\,Hz was added for
each realisation with trial glitch epoch moving across the detection
window at step 30\,d. We introduce $\sigma_{\rm ep}$, absolute
difference between returned glitch epoch and input epoch over average
ToA interval, and relative size error $\epsilon_{\Delta\nu}$, absolute
difference between returned glitch size and input size over input
size, to characterise the accuracy of the returned values
respectively. Results showed $\sigma_{\rm ep}$ values were consistent
with zero to at least seven decimal places, $\epsilon_{\Delta\nu}$
values were consistent with zero to at least six decimal places. Then,
in test two, we moved toward reality by introducing real observing
sampling (the observed, refined integer phases) and ToA uncertainties
into the simulation to timing residuals. A glitch also with
$\Delta\nu=10^{-7}$\,Hz was moved across the detection window by an
average ToA interval (23\,d) over realisations. Results showed both
$\sigma_{\rm ep}$ and $\epsilon_{\Delta\nu}$ values were consistent
with zero to at least two decimal places. Then, in test three, we
further added timing noise to the simulated timing residuals. For the
spectral parameters (see equation \eqref{eq:pden}), we adopted $f_{\rm
  c}=0.06$\,yr$^{-1}$, the reciprocal of the observing time span,
$\alpha=4.0$, the limit of the steepness for first order pre-whitening
to overcome `spectral leakage' \citep{chc+11}, and
$A=1.0\times10^3$\,s$^3$, an arbitrary strength to which a
$\Delta\nu=10^{-7}$\,Hz glitch is well identifiable by eye. In Figure
\ref{fig:edge}, we plot the variations of $\sigma_{\rm ep}$ and
$\epsilon_{\Delta\nu}$ with respect to the fraction of trial glitch
epoch within the detection window. As another check, we, in test four,
changed $A$ to $1.0\times10^8$\,s$^3$ and $\Delta\nu$ to
$10^{-5}$\,Hz. Results are also plotted in Figure \ref{fig:edge}. The
largest $\sigma_{\rm ep} \sim 1.9$ was found in test four but it is
still smaller than 3.0, the criterion for resolving two glitches. Most
$\sigma_{\rm ep}$ values are consistent with zero to two decimal
places. For the largest $\epsilon_{\Delta\nu} \sim 0.75$ occurred in
test three, the size returned is away within a factor of two. Most
$\epsilon_{\Delta\nu}$ values are consistent with zero to two decimal
places.

\begin{figure}
\begin{center}
\includegraphics[angle=-90,width=8cm]{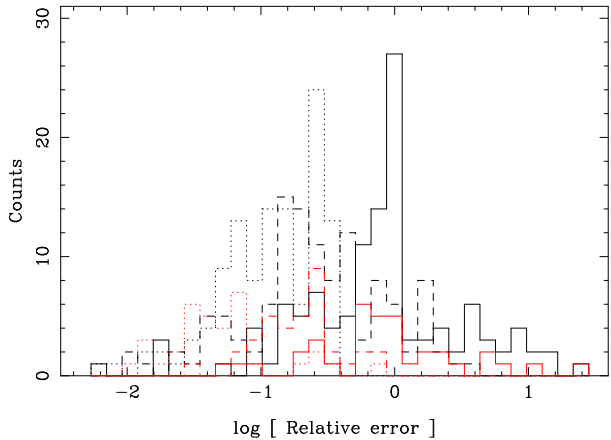}
\end{center}
\caption{Histogram (32 bins) of spectral parameters' relative errors
  which are from the examination to \textsc{temponest} red noise
  modelling. Solid bars are for $A$, the amplitude of power spectral
  density. Dashed bars are for $f_{\rm c}$, the spectral corner
  frequency. Dotted bars are for $\alpha$, the spectral
  exponent. Black is for the 121 pulsars that were not seen to glitch
  in the \citeauthor{ymh+13} sample, while red is for the observed 36
  glitching pulsars.}\label{fig:rnexam}
\end{figure}

The other point we ought to examine is how well \textsc{temponest}
models red noise. This is essentially required before we
determine ${\rm P}\{C_{\rm noise}\}$. In this examination, we no
longer involved only an individual pulsar but enlarged our sample by
including as many as \citeauthor{ymh+13} pulsars if phase-connected
timing solutions could be obtained over entire data spans. We thus
involved 157 pulsars, all 36 observed glitching pulsars were
included. For the other eight, there commonly exist large data gaps
for typically thousands of days (and overall timing solutions were not
obtained)\footnote{The eight pulsars are PSRs~J1016$-$5819,
  J1327$-$6400, J1524$-$5625, J1541$-$5535, J1637$-$4642,
  J1821$-$1419, J1853$-$0004 and J1853$+$0011.}. Our scheme for this
examination was to model simulated red noise for each of the 157
pulsars. In the simulations, real observing sampling and ToA
uncertainties were used. For the input spectral parameters, as in last
examination, we took the reciprocal of the observing time spans for
corner frequencies, and we fixed exponents at 4.0 for all pulsars. But
we no longer arbitrarily set noise strength. For a given pulsar with
timing residuals $r_i$ ($i=0,1,$\ldots$,n$), we roughly
estimated the amplitude of the power spectral density $A$ as
\begin{equation}
\frac{T}{n^2} \sum_i r_i^2 \sim \frac{T}{n} {\rm rms}^2 \sim
T\cdot{\rm rms}^2, \label{eq:aesti}
\end{equation}
or the product of the observing time span and the mean-square
residual. In the equation, the summation over $r_i$ squares is the
standard deviation of the power spectrum of a white noise sequence
\citep{rem02}. (So we call this estimate `rough'.)  As in last
examination, we use relative error to characterise the accuracy of
returned values. We did one realisation for each pulsar. Figure
\ref{fig:rnexam} shows the histogram of the relative errors of the
spectral parameters. We found, out of the $157\times3$ measurements,
412 (87.5\%) had a relative error smaller than 1.0. In particular, for
the measurements of $A$, this proportion was 114/157. After the
examinations, we moved on to fit red noises for real data.
\section{Solution}\label{sec:solun}
\subsection{Noise term}\label{sec:solun:term_n}
\begin{figure}
\begin{center}
\includegraphics[angle=-90,width=8cm]{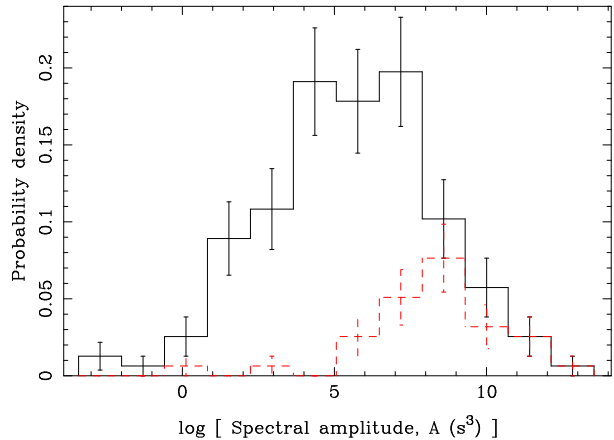}
\end{center}
\caption{Probability density of the amplitude of power spectral
  density of 157 pulsars in the Yu et al. sample, ${\rm P}\{C_{\rm
    noise}\}$ (12 bins). Red dashed bars indicate contribution of the
  36 observed glitching pulsars.}\label{fig:p_n}
\end{figure}

\begin{figure*}
\begin{center}
\begin{tabular}{cc}
\includegraphics[angle=-90,width=8cm]{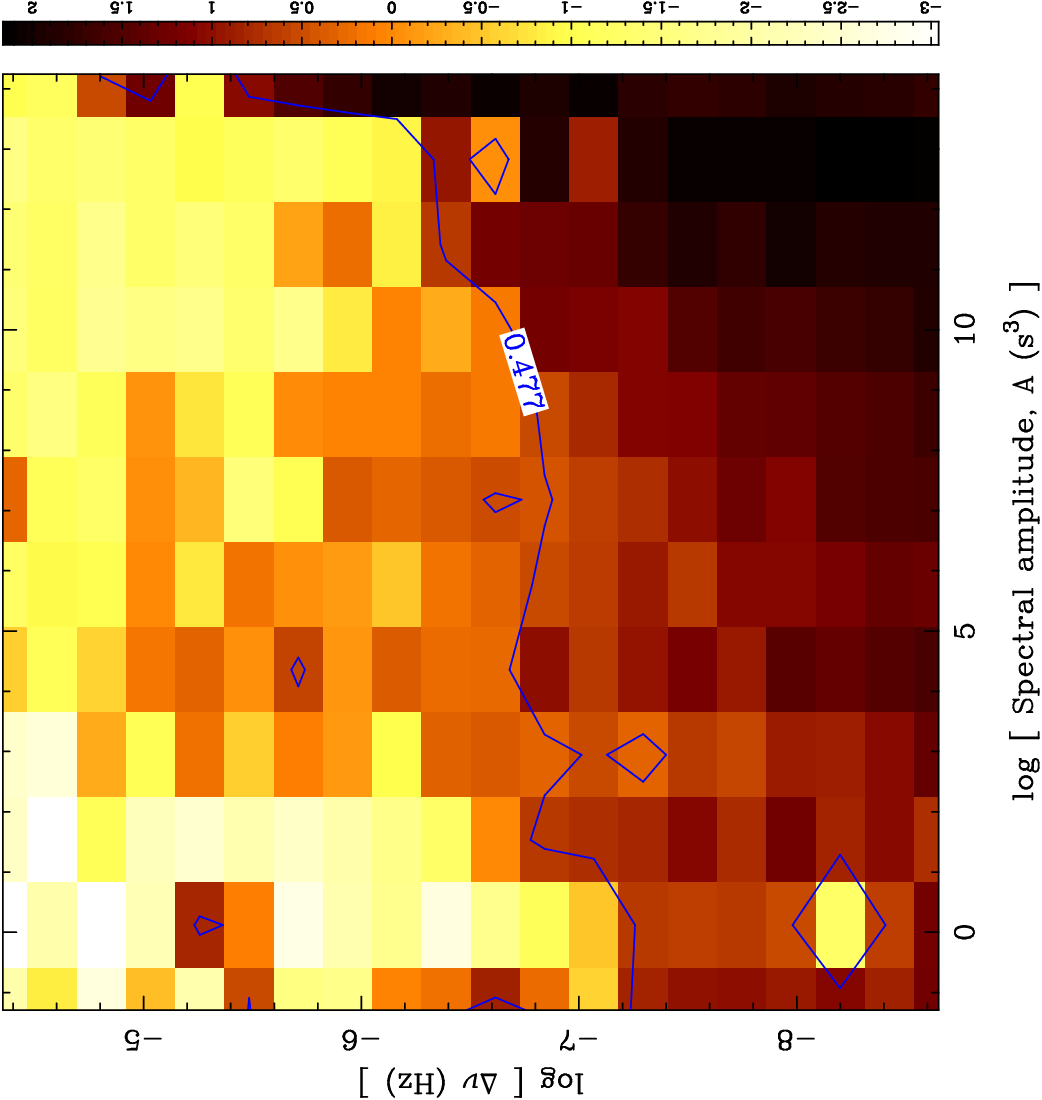} & \includegraphics[angle=-90,width=8cm]{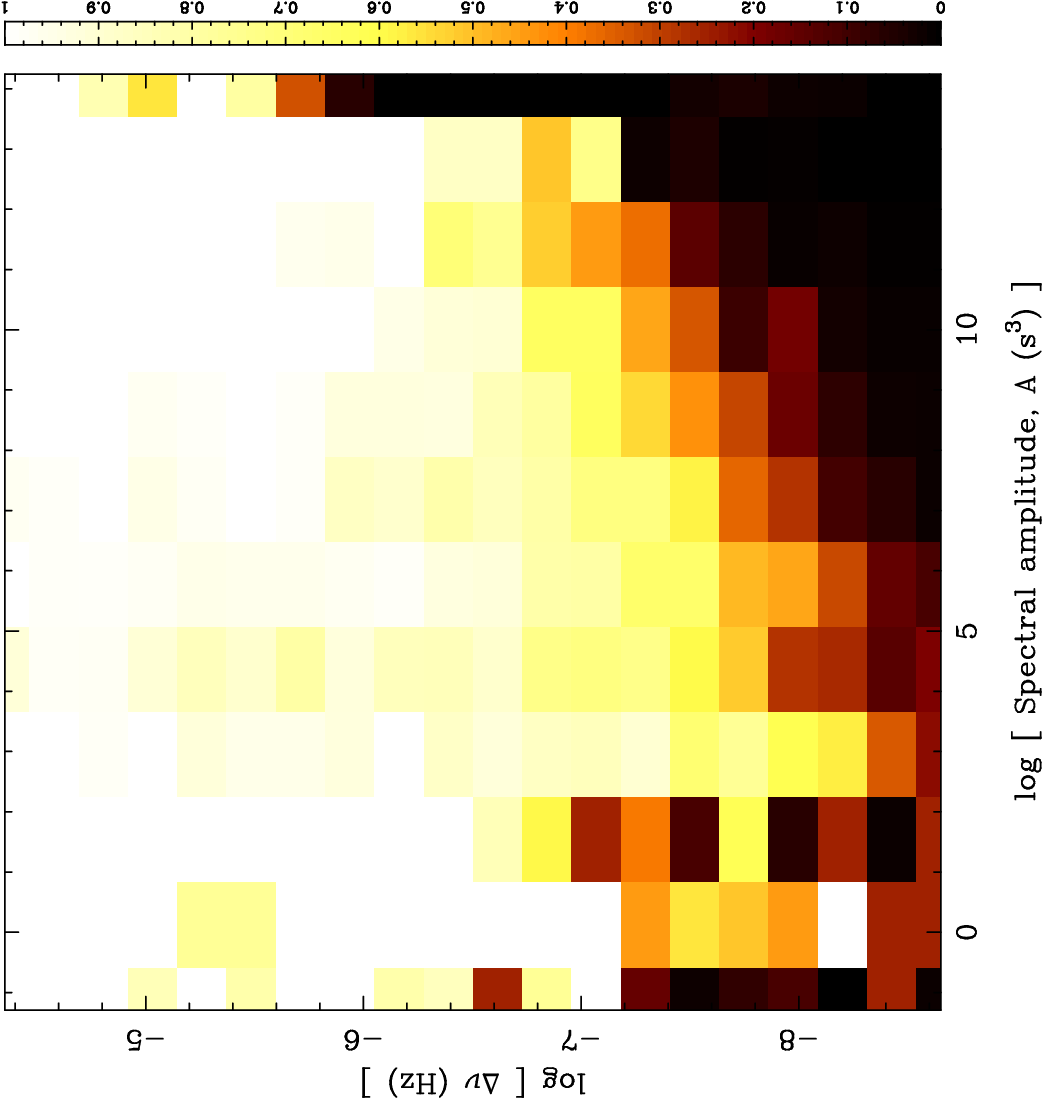}\\
\end{tabular}
\end{center}
\caption{Left panel: The $12\times20$ grid showing $\sigma_{\rm ep}$
  variations as a function of timing noise strength and glitch
  size. The contour is at $\sigma_{\rm ep}=3.0$. Pixel values are in
  logarithmic scale. Right panel: Same grid showing the solution to
  the ${\rm P}\{D(\Delta\nu)|C_{\rm noise}\}$ matrix. Each cell
  indicates the ratio of the number of positive detections to the
  total number of glitches distributed into the cell. To present the
  image clearer, value for each pixel is squared.}\label{fig:imag}
\end{figure*}

With the awareness of the performance of \textsc{temponest} red noise
modelling, we fitted red noises for the 157 pulsars. In Figure
\ref{fig:p_n}, the top panel shows ${\rm P}\{C_{\rm noise}\}$, the
probability density of the amplitude of power spectral density $A$,
derived by normalising the various counts to the binned logarithmic
amplitude values. Uncertainties were determined as the square root of
the counts followed by the same normalisation. As our sample is not
large, the bin number twelve is the largest that avoids a void bin. It
shows that high end of the distribution is more contributed by the
pulsars that have been observed to glitch (than by the pulsars
otherwise). This could be because unmodelled glitch features have
caused red noises.

After we measured spectral parameters, we were able to run the Monte
Carlo simulation to determine ${\rm P}\{D(\Delta\nu)|C_{\rm
  noise}\}$. In total, we made 100 realisations for each pulsar. In
each of the realisations, a glitch event with an epoch uniformly
distributed within the specific detection window and a size uniformly
distributed between $1.65\times10^{-9}$ and $3.52\times10^{-5}$\,Hz
(exclusive) was added into the simulated timing residuals; size
boundaries were defined by the minimum and maximum detected sizes in
the \citeauthor{ymh+13} data. Before adding a glitch, real pulsar
ephemeris, observing sampling, ToA uncertainties and the measured
spectral parameters were used to produce residuals. Although more
realisations might be desirable, we were restricted by the expensive
computation of running the \textsc{multinest} at
\texttt{double-double} precision. The option of the high precision was
to robustly calculate the Bayesian evidence. Its value, for some
cases, has expanded up to a few thousand in logarithmic scale. In
practice, such an integration has required an
Intel$^{\textregistered}$ 2.5GHz processor to take hours to complete
and, for model evaluation, tens of sampling (to the parameter space)
were typically made. Difficulties in the evidence evaluation have been
fully interpreted by \citet{fhb09}.

As in Section \ref{sec:routn:exam}, $\sigma_{\rm ep}$ can be the index
for glitch location. We scatter the obtained 15,700 $\sigma_{\rm ep}$
values onto a plane. One dimension denotes strength of timing noise,
one dimension denotes size of glitches. To illustrate, we average the
values scattered into the same cell, as shown in Figure \ref{fig:imag}
left panel. The twelve-by-twenty grid was chosen such that none of the
cells is empty. It is natural to see that it is easier to detect
glitches with larger sizes in weaker timing noises. We then draw a
contour at 3.0 to illustrate the criterion for `positive'
detections. In other words, we define those detections with
$\sigma_{\rm ep}<3.0$ as `positive', the others are `negative'. Under
this definition, we derive the detection probability for each cell on
this plane (${\rm P}\{D(\Delta\nu)|C_{\rm noise}\}$) as the ratio of
the number of positives to the number of the total scattered into the
cell. In Figure \ref{fig:imag}, the right panel illustrates this
matrix. In fact, this figure was expected to present the same pattern
as the left panel. For each element in the matrix, uncertainty was
determined via square root of the counts and error propagation. Now,
we do product for the matrices ${\rm P}\{D(\Delta\nu)|C_{\rm noise}\}$
and ${\rm P}\{C_{\rm noise}\}$ followed by a normalisation to derive
solution of the noise term. Result is given in Figure \ref{fig:p_tot}.
\subsection{Epoch term and multi-glitch term}\label{sec:solun:term_e_m}
\begin{figure}
\begin{center}
\includegraphics[angle=-90,width=8cm]{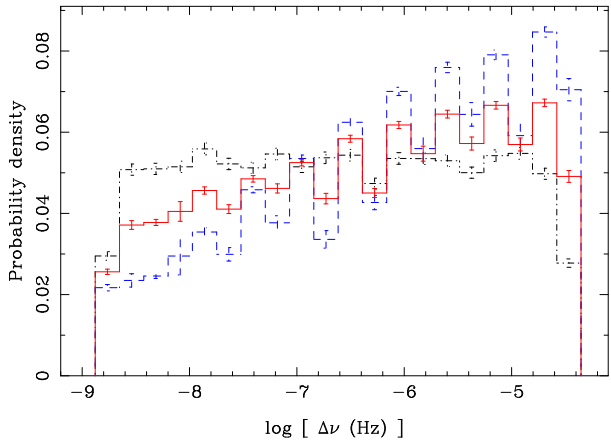}
\end{center}
\caption{Solution of glitch detection probability as a function of
  glitch size presented in the form of probability density (20
  bins). Blue dashed bars indicate the noise term, black dashed-dotted
  bars indicate the epoch and multi-glitch terms and red solid bars
  indicate the complete probability.}\label{fig:p_tot}
\end{figure}

After solving for the noise term, we implemented equations
\eqref{eq:p_e} to \eqref{eq:p_d|m} to derive ${\rm
  P}\{D(\Delta\nu)|C_{\rm epoch}\}{\rm P}\{C_{\rm epoch}\} + {\rm
  P}\{D(\Delta\nu)|C_{\rm multi}\}{\rm P}\{C_{\rm multi}\}$ for each
of the glitches simulated. Then we scattered the values into the
twenty $\Delta\nu$ bins, made sum in each bin and did
normalisation. Uncertainty for each bin was determined as the square
root of the bin value followed by the same normalisation. It had been
expected that the distribution would be uniform as we uniformly
generated $\Delta\nu$ values and scattered them uniformly. Result
plotted in Figure \ref{fig:p_tot} confirms this. The low at the
boundaries reflects the fact that we generated $\Delta\nu$ values in
the open interval. Finally, we added the partial detection probability
densities of the epoch, multi-glitch and noise terms in each
$\Delta\nu$ bin together and made normalisation to obtain solution for
the complete probability formula (equation
\eqref{eq:prob}). Uncertainty for each bin was determined by error
propagation and the same normalisation. Result is plotted in Figure
\ref{fig:p_tot}. We see that detectability of glitches is not uniform
with respect to glitch sizes, the detection probability density
becomes more and more massive as glitch becomes large. Fluctuations
result from the finite scale of our simulation.

\section{Discussion}\label{sec:disn}
\subsection{Aggregated distribution}\label{sec:disn:adistr}
\begin{figure}
\begin{center}
\includegraphics[angle=-90,width=8cm]{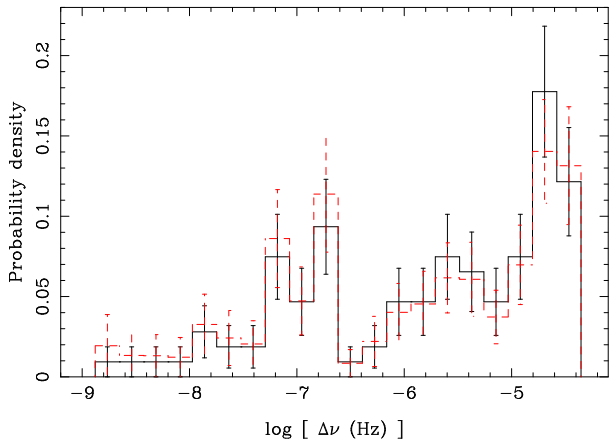}
\end{center}
\caption{The apparent (black solid) and the inferred (red dashed)
  aggregated glitch size distribution.}\label{fig:prob_dv}
\end{figure}

With the detection probability densities derived, we are able to infer
the aggregated glitch size distribution embedded in the
\citeauthor{ymh+13} data sets. In the histograms in Figure
\ref{fig:prob_dv}, the solid bars present the observed
distribution. We divide the observations by the detection probability
densities. After a normalisation, we present the result with the
dashed bars. We see that the inferred distribution appears similar to
the observed distribution. This means most glitches embedded in the
data sets are detectable. Using the manual searching method
\citeauthor{ymh+13} have detected the glitches that could be
detected. The similarity also means we have set up a good model for
glitch detectability of the data sets and manual method. However, when
we observe the distributions more carefully, we find the inferred
distribution becomes a bit more massive than the observed one for
glitches with $\Delta\nu \lesssim 4\times10^{-7}$\,Hz, implying
\citeauthor{ymh+13} were unable to detect some small glitches. This
would result from the lower detectability for small glitches of the
data sets. To study this in more detail, next, we shall determine the
detection probability densities for each of the pulsars
PSRs~J1048$-$5832, J1341$-$6220, J1413$-$6141, J1420$-$6048,
J1740$-$3015, J1801$-$2304 and J1801$-$2451; they are the pulsars that
have glitch numbers $\geqslant 5$ in the data.
\subsection{Individual distributions}\label{sec:disn:idistr}
\begin{figure}
\begin{center}
\begin{tabular}{c}
\includegraphics[angle=-90,width=8cm]{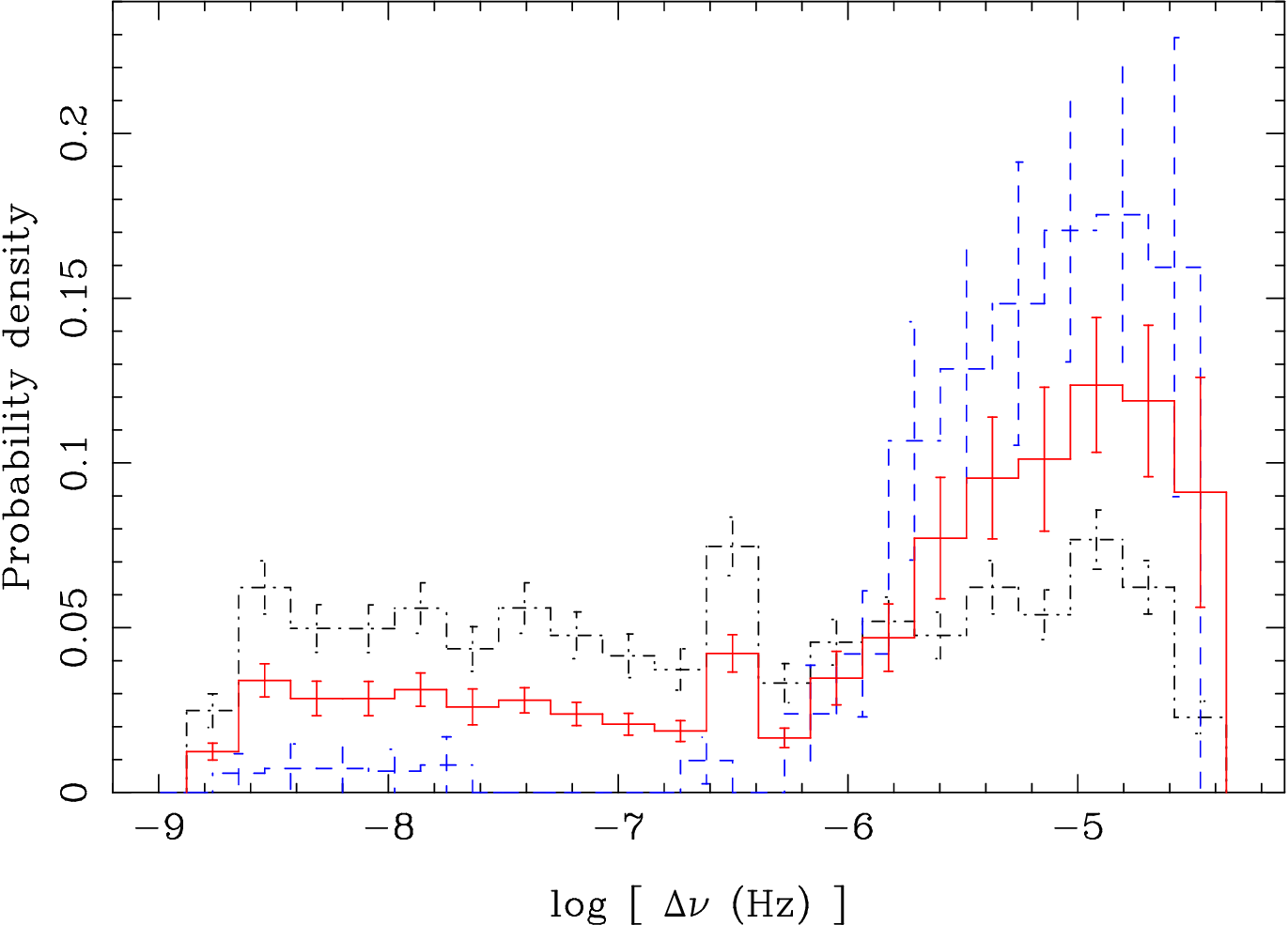}\\
\includegraphics[angle=-90,width=8cm]{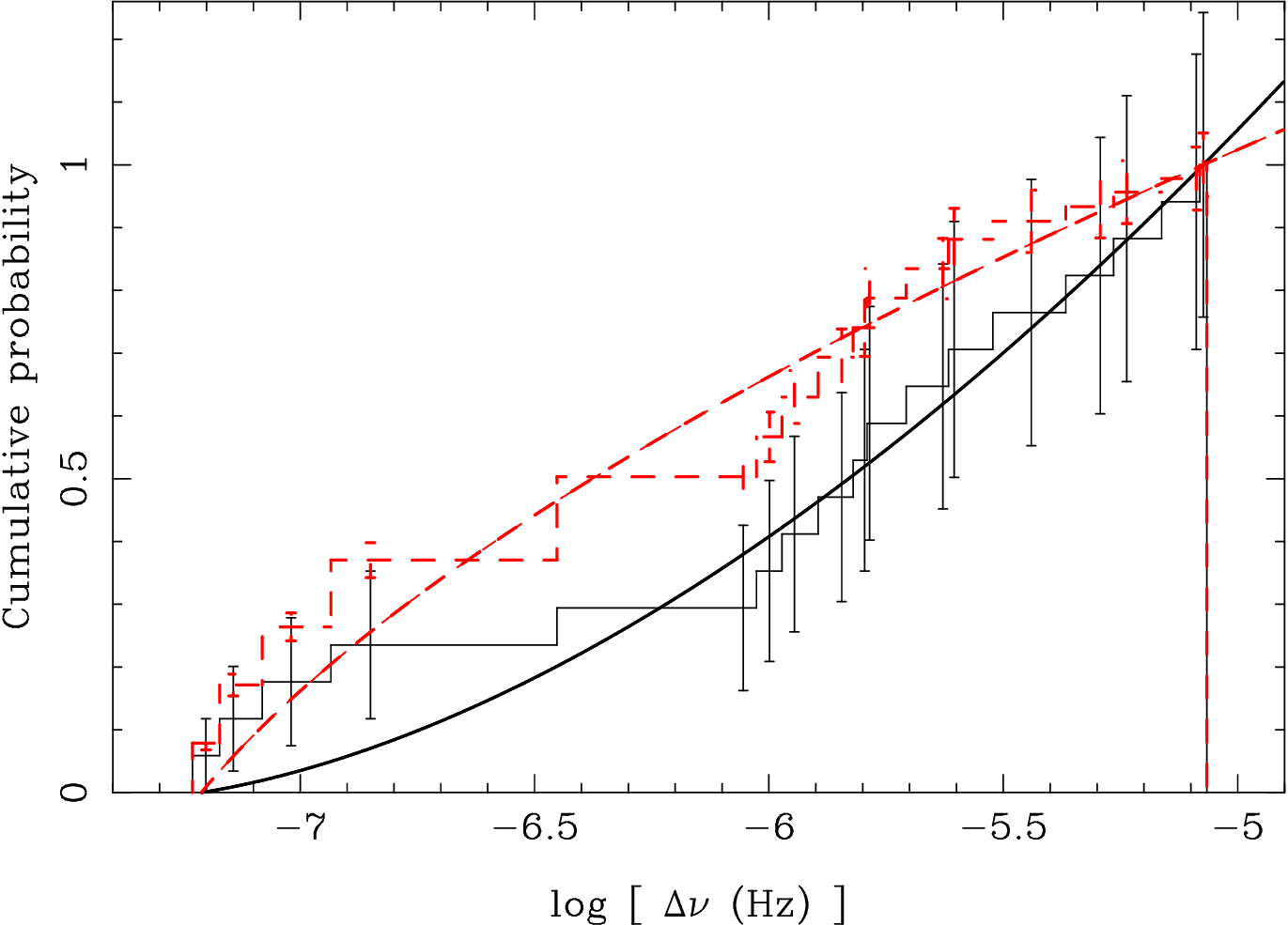}
\end{tabular}
\end{center}
\caption{Upper panel: Glitch detection probability densities of
  PSR~J1341$-$6220. Blue dashed bars indicate the noise term, black
  dashed-dotted bars indicate the epoch and multi-glitch terms and red
  solid bars indicate the complete probability. Binning is the same as
  Figure \ref{fig:p_tot}. To present the histograms clearer, the black
  dashed-dotted and red solid bars are shifted right by half bin
  size. Lower panel: The observed (black solid) and inferred (red
  dashed) cumulative distribution of glitch sizes for
  PSR~J1341$-$6220. Curves are the fitted power-law models (see
  text).}\label{fig:idistr_J1341}
\end{figure}

Among the seven pulsars PSR~J1341$-$6220 was observed to show
seventeen glitches from MJD 49540 to MJD 55461, presenting the largest
glitch number and highest glitching rate. We study it first. Similar
to the group study we used the Monte Carlo method to solve the
complete probability formula (equation \ref{eq:prob}). We recall the
description in Section \ref{sec:prob} that the only difference of the
probability definition for individual case from group case is the
${\rm P}\{C_{\rm noise}\}$ piece is an arbitrary number rather than a
distribution. We thus made it one. In each realisation, real pulsar
ephemeris, observing sampling, ToA uncertainties and the measured
power spectral parameters were used to generate timing residuals; a
glitch with epoch uniformly distributed within the detection window
and size uniformly distributed between $1.65\times10^{-9}$ and
$3.52\times10^{-5}$\,Hz (exclusive) was then added into the
simulation. \textsc{temponest} was used for the glitch search,
parameter space was defined in the same way as in the group
study. Detections satisfying $\sigma_{\rm ep} < 3.0$ were recognised
as positive. Up until the final preparation for this section, we
accumulated 482 realisations. In Figure \ref{fig:idistr_J1341} the
upper panel gives the derived detection probability densities;
uncertainties were determined in the same way as in the group
study. We see that the densities vary around 0.02 up until $\Delta\nu
\sim 10^{-6}$\,Hz, then the densities grow to around 0.1. In other
words the probability for detecting a glitch with $\Delta\nu \lesssim
10^{-6}$\,Hz is about three times smaller than the probability for
detecting a glitch with $\Delta\nu \gtrsim 10^{-6}$\,Hz. This would
cause bias in our knowledge of the pulsar's glitch size
distribution. In Figure \ref{fig:idistr_J1341} lower panel the solid
bars present cumulative distribution function (CDF) of the glitch
sizes observed in PSR~J1341$-$6220; uncertainty for each bin was
determined as square root of the count followed by the same
normalisation. We least-squares modelled the CDF using the function
\begin{equation}
P(<\Delta\nu) = \frac{\Delta\nu^{1+s}-\Delta\nu_{\rm
    min}^{1+s}}{\Delta\nu_{\rm max}^{1+s}-\Delta\nu_{\rm
    min}^{1+s}}\label{eq:cdf_pl}
\end{equation}
with fixing $\Delta\nu_{\rm min}$ and $\Delta\nu_{\rm max}$ at the
minimum and maximum glitch sizes observed in the pulsar
respectively. We obtained $s=0.7^{+1.4}_{-0.7}$. The following
Kolmogorov-Smirnov (K-S) test gave $Q_{KS}=0.20$ the probability that
the null hypothesis, the data and model are drawn from the same
distribution, is false. Uncertainties for $s$ were determined as the
boundaries of 68 per cent confidence level to reject the null
hypothesis. To infer the glitch size distribution embedded in the data
with the observed distribution and derived detection probability
densities, we first binned the observed $\Delta\nu$ values into the
twenty density bins; for bin $i$, count is denoted $m_i$. Then we
picked up those bins with $m_i>0$. When forming the inferred CDF we
used the nearest integer of $m_i/p_i$ ($p_i$ is the density value of
bin $i$) instead of one as the step at each $\Delta\nu$
value. Uncertainty for each bin was determined via error
propagation. Result is plotted in Figure \ref{fig:idistr_J1341} lower
panel as dashed bars. K-S test gave $Q_{KS}=0.10$ when comparing the
data with the model with an exponent $s=-0.4^{+1.0}_{-0.4}$. This
verifies our argument that the low detectability for small glitches of
the data has biased our knowledge of PSR~J1341$-$6220's glitch size
distribution.

\begin{figure*}
\begin{center}
\begin{tabular}{cc}
\includegraphics[angle=-90,width=7cm]{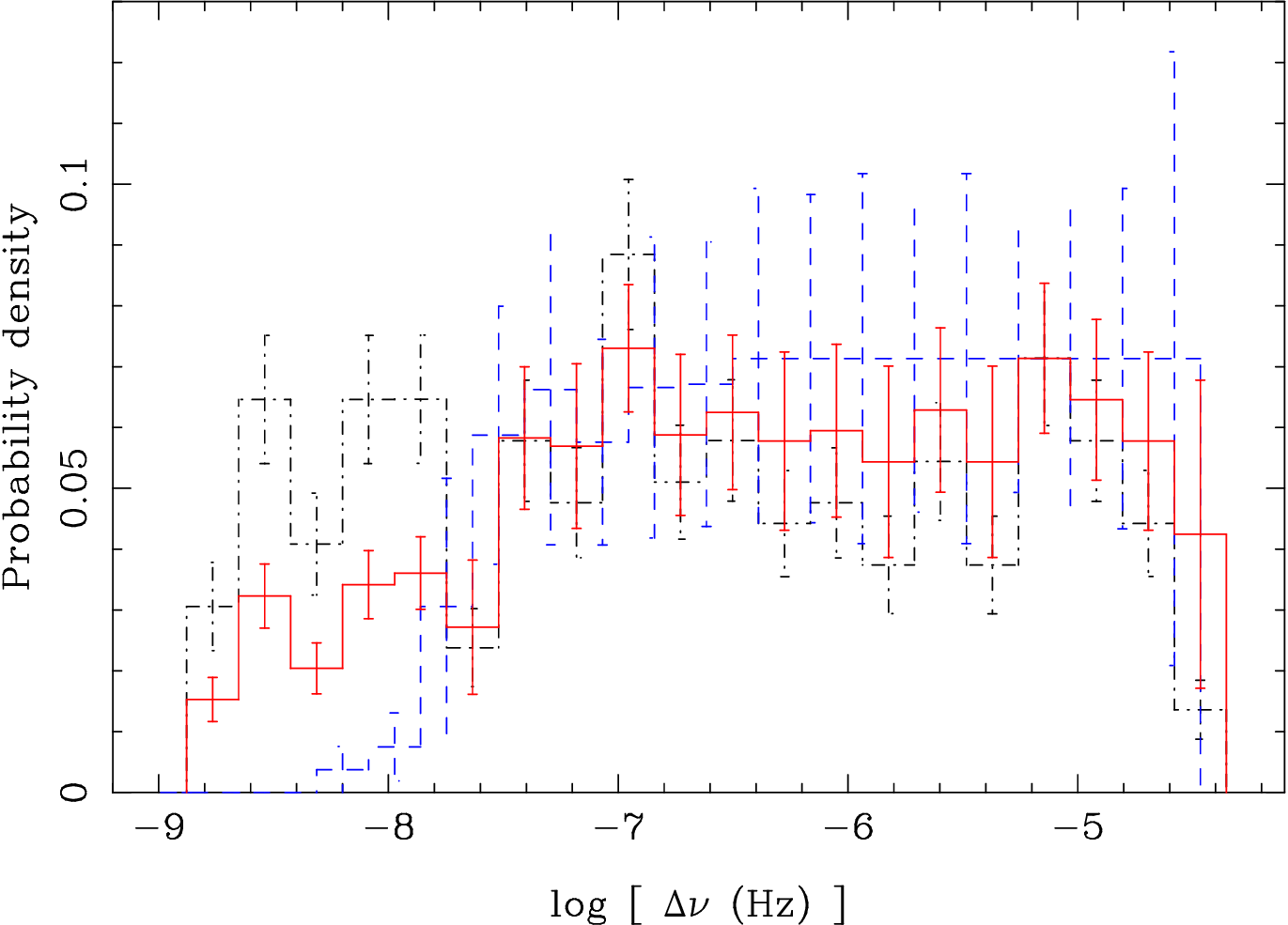} &
\includegraphics[angle=-90,width=7cm]{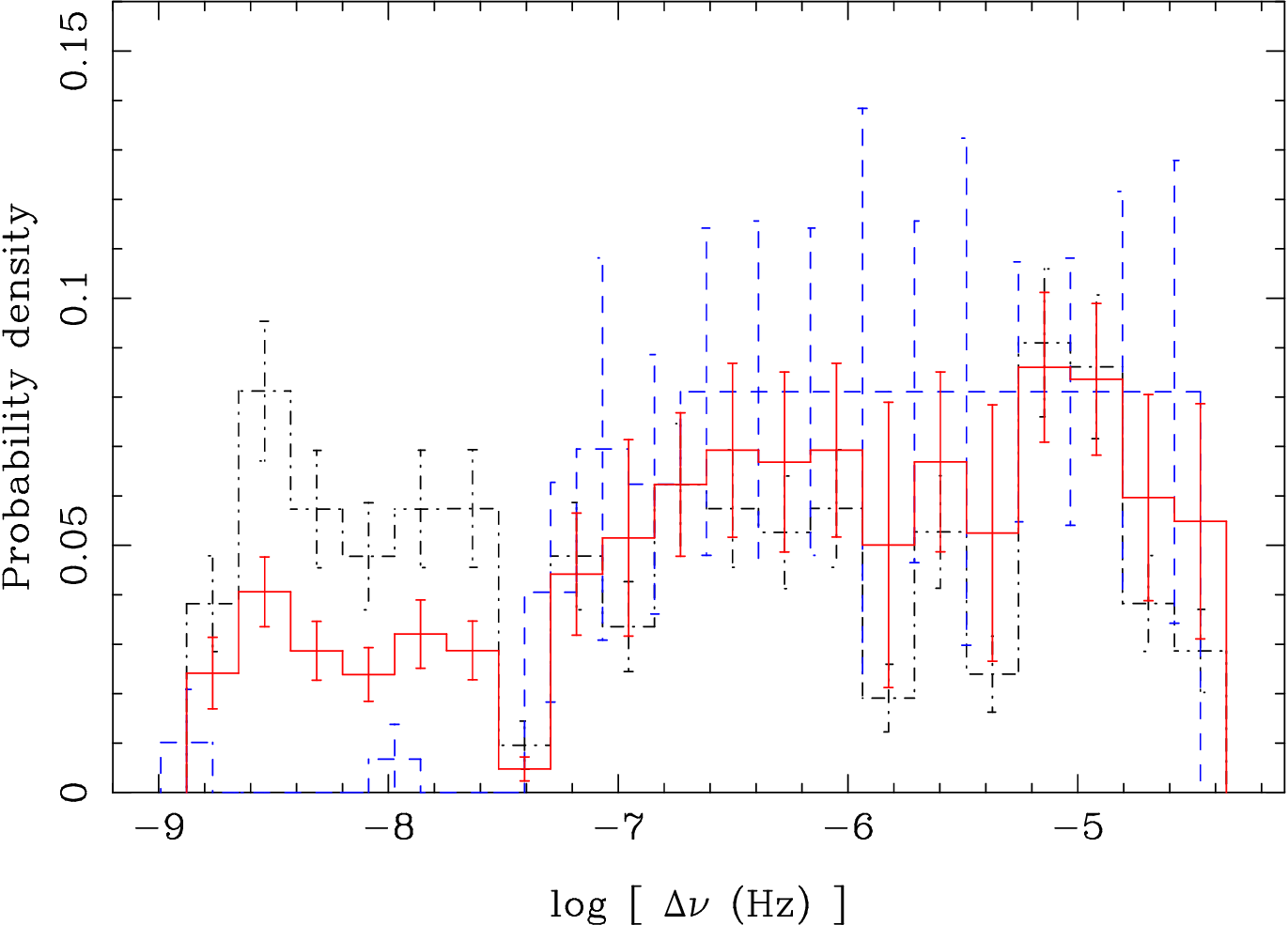}\\
\includegraphics[angle=-90,width=7cm]{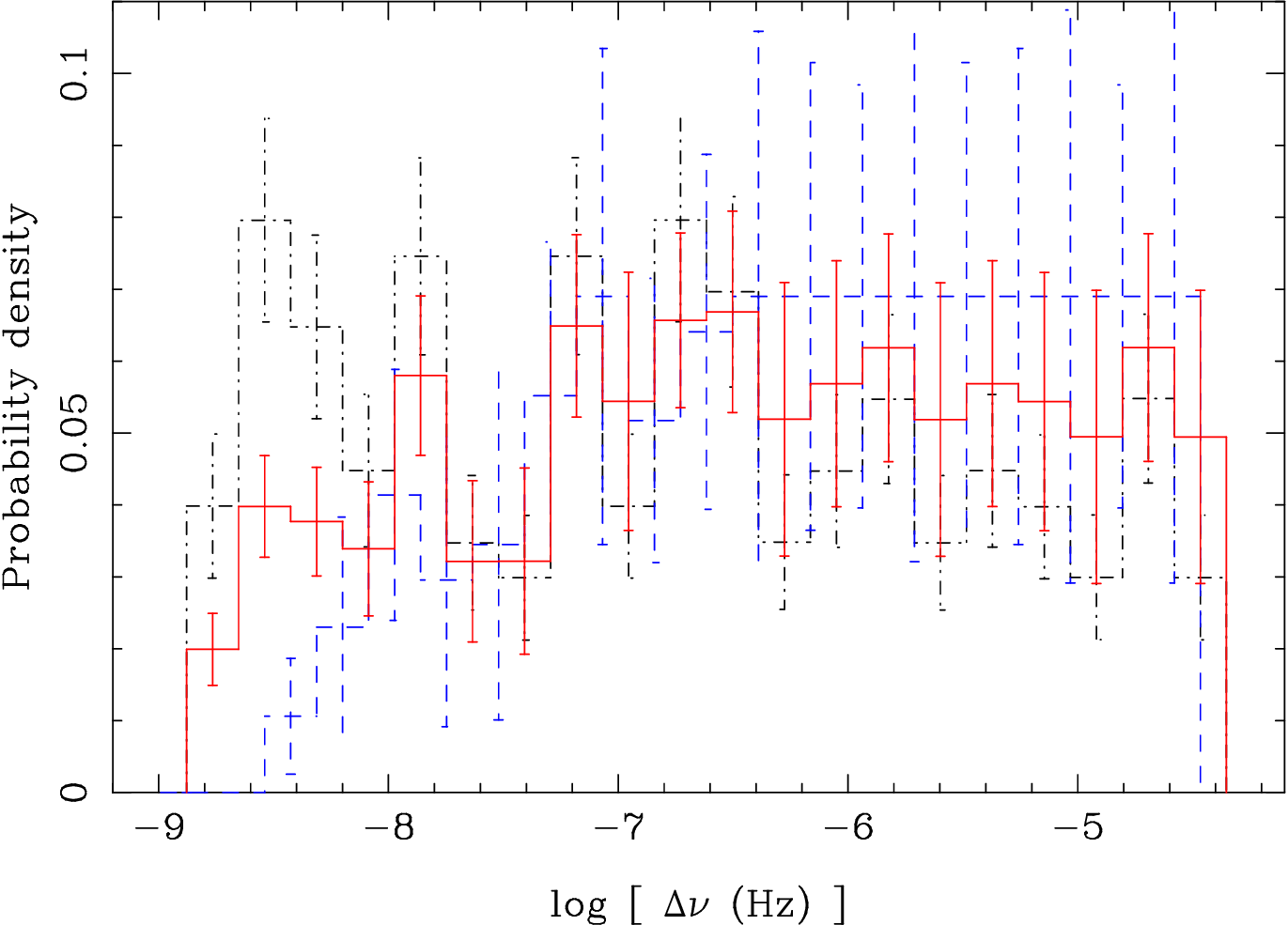} &
\includegraphics[angle=-90,width=7cm]{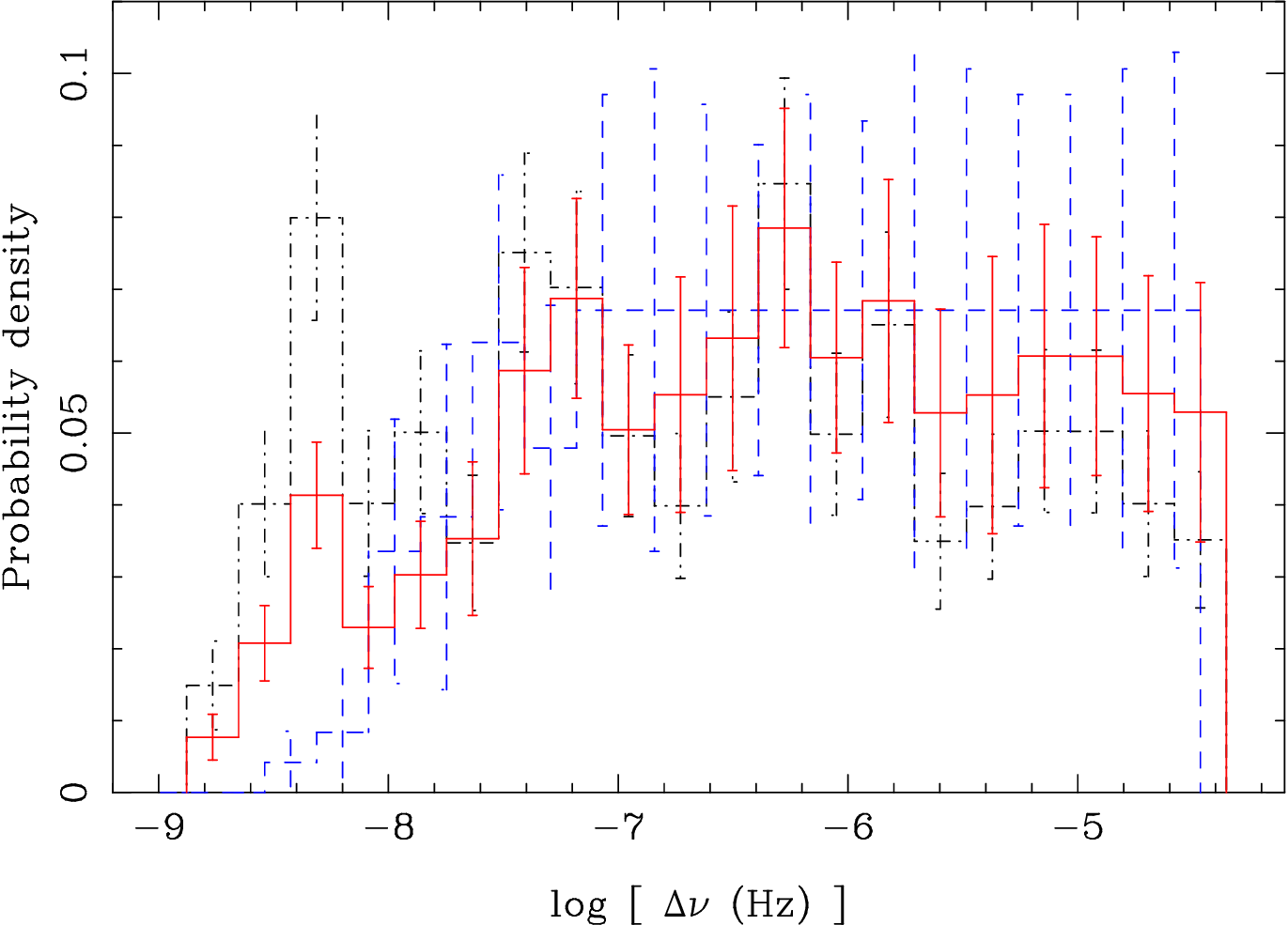}\\
\includegraphics[angle=-90,width=7cm]{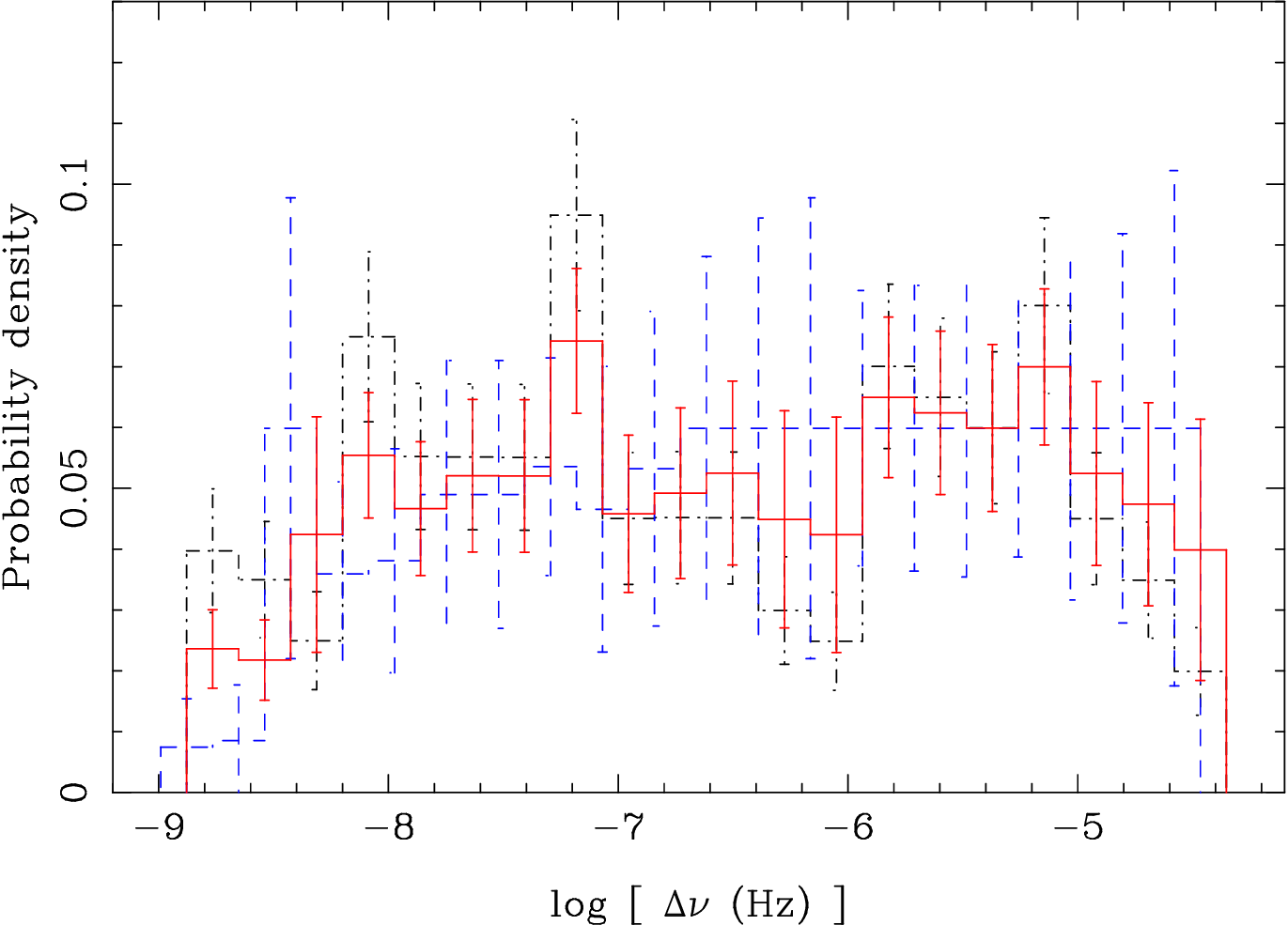} &
\includegraphics[angle=-90,width=7cm]{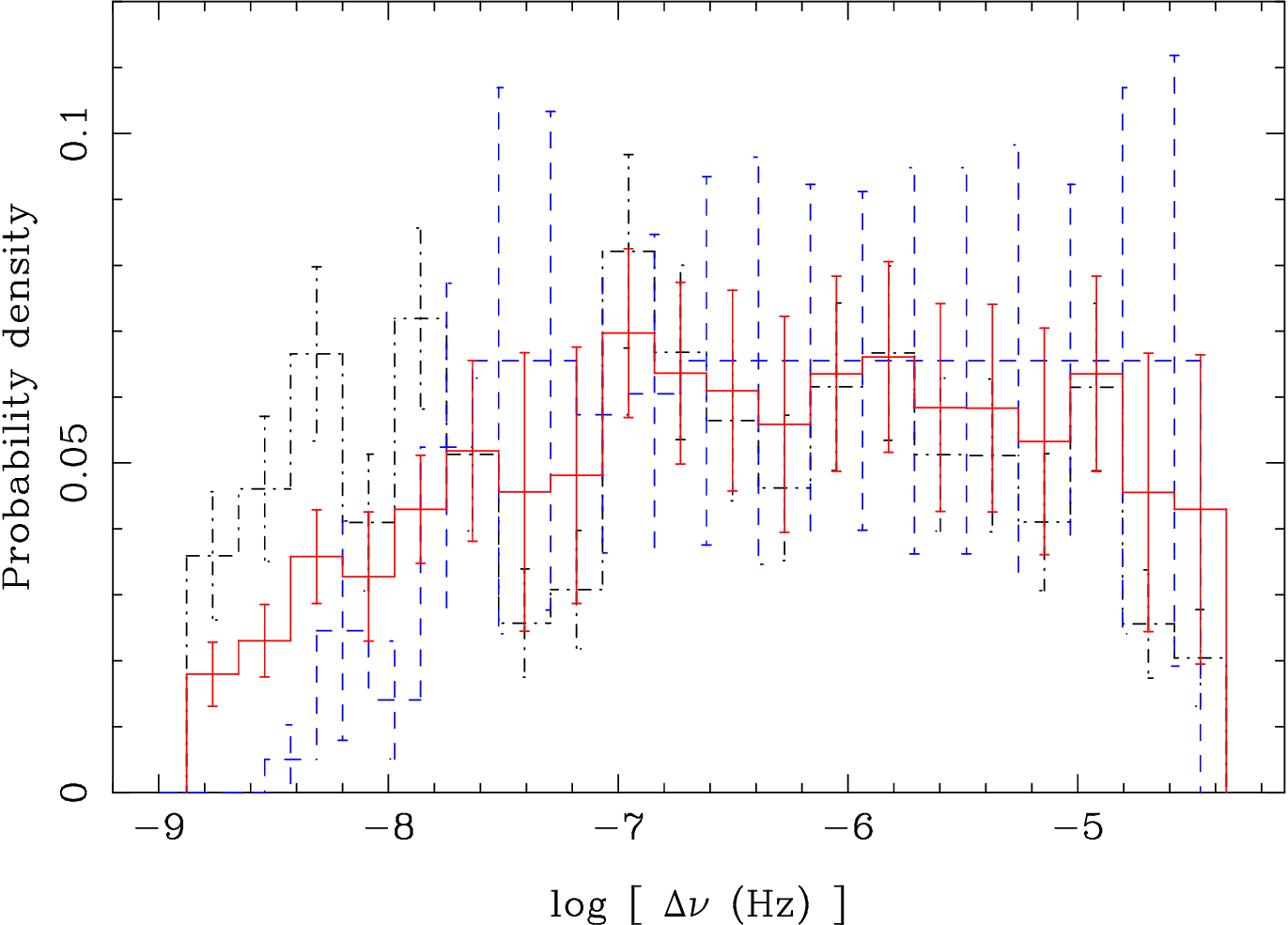}\\
\end{tabular}
\end{center}
\caption{Glitch detection probability densities of PSRs~J1048$-$5832
  and J1413$-$6141 (first row), PSRs~J1420$-$6048 and J1740$-$3015
  (second row), PSRs~J1801$-$2304 and J1801$-$2451 (third row). Blue
  dashed bars indicate the noise term, black dashed-dotted bars
  indicate the epoch and multi-glitch terms and red solid bars
  indicate the complete probability. Binning is the same as Figure
  \ref{fig:p_tot}. To present the histograms clearer, the black
  dashed-dotted and red solid bars are shifted right by half bin
  size.}\label{fig:idistr_p_tot}
\end{figure*}

\begin{figure*}
\begin{center}
\begin{tabular}{cc}
\includegraphics[angle=-90,width=7cm]{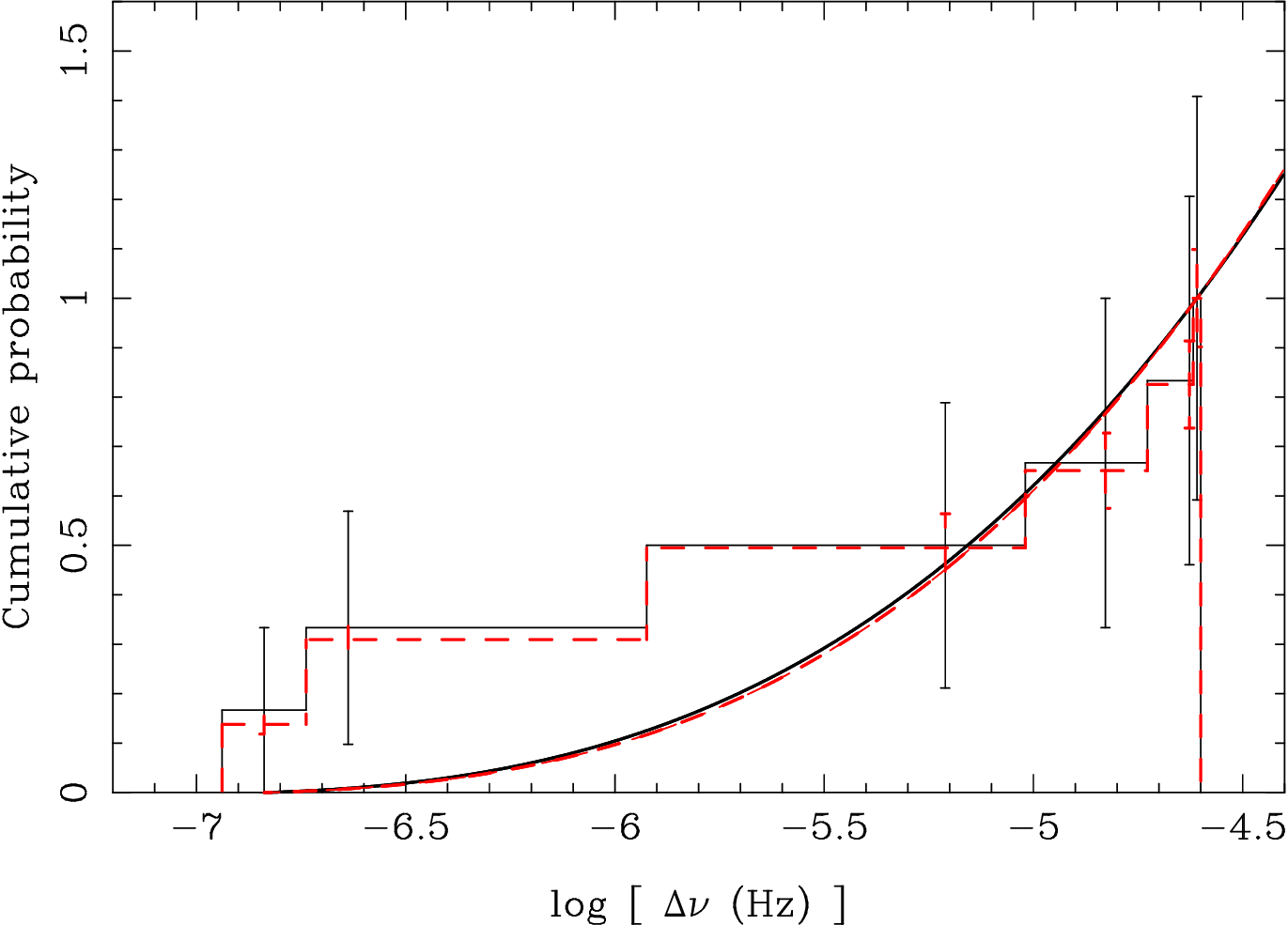} &
\includegraphics[angle=-90,width=7cm]{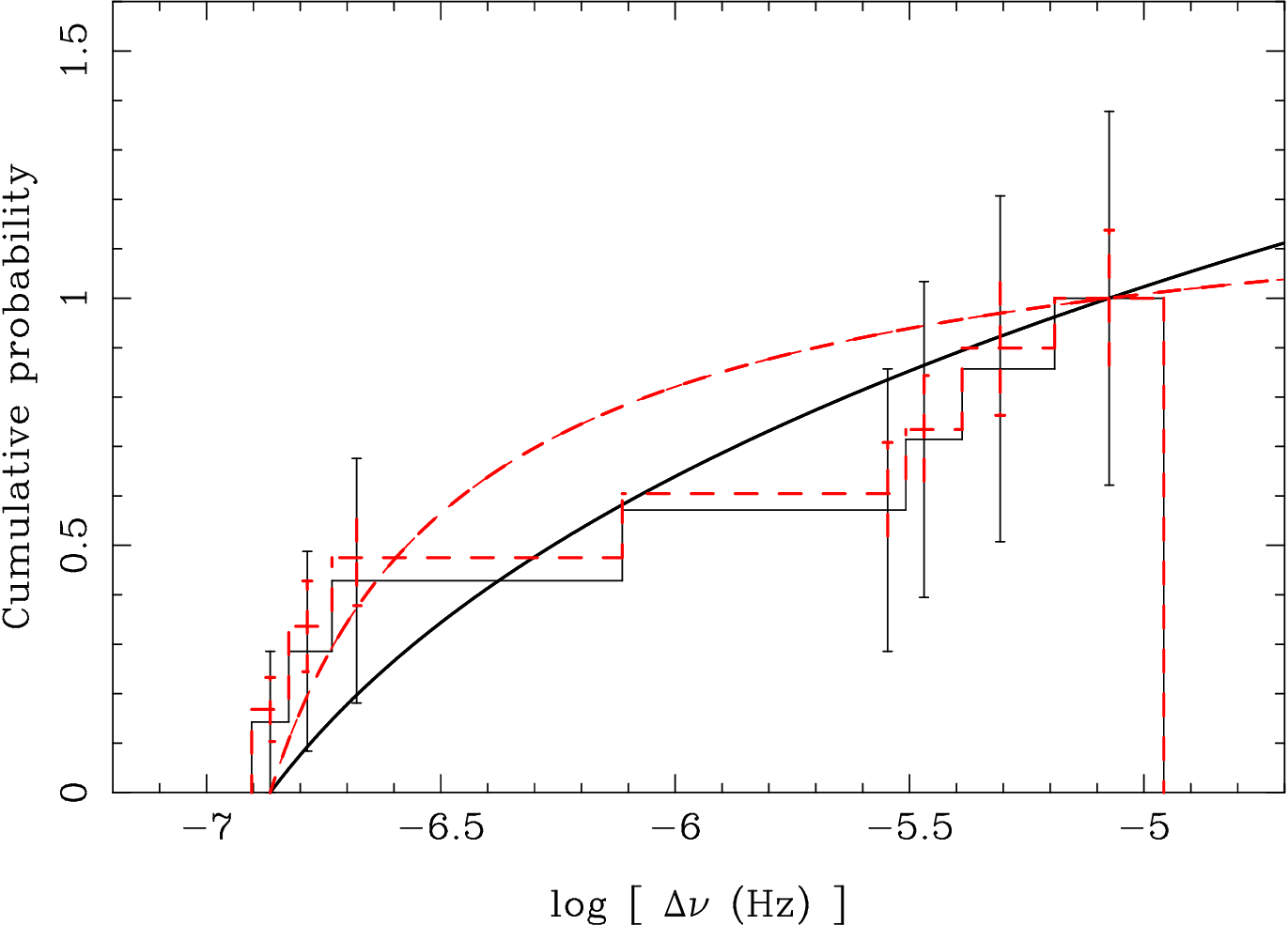}\\
\includegraphics[angle=-90,width=7cm]{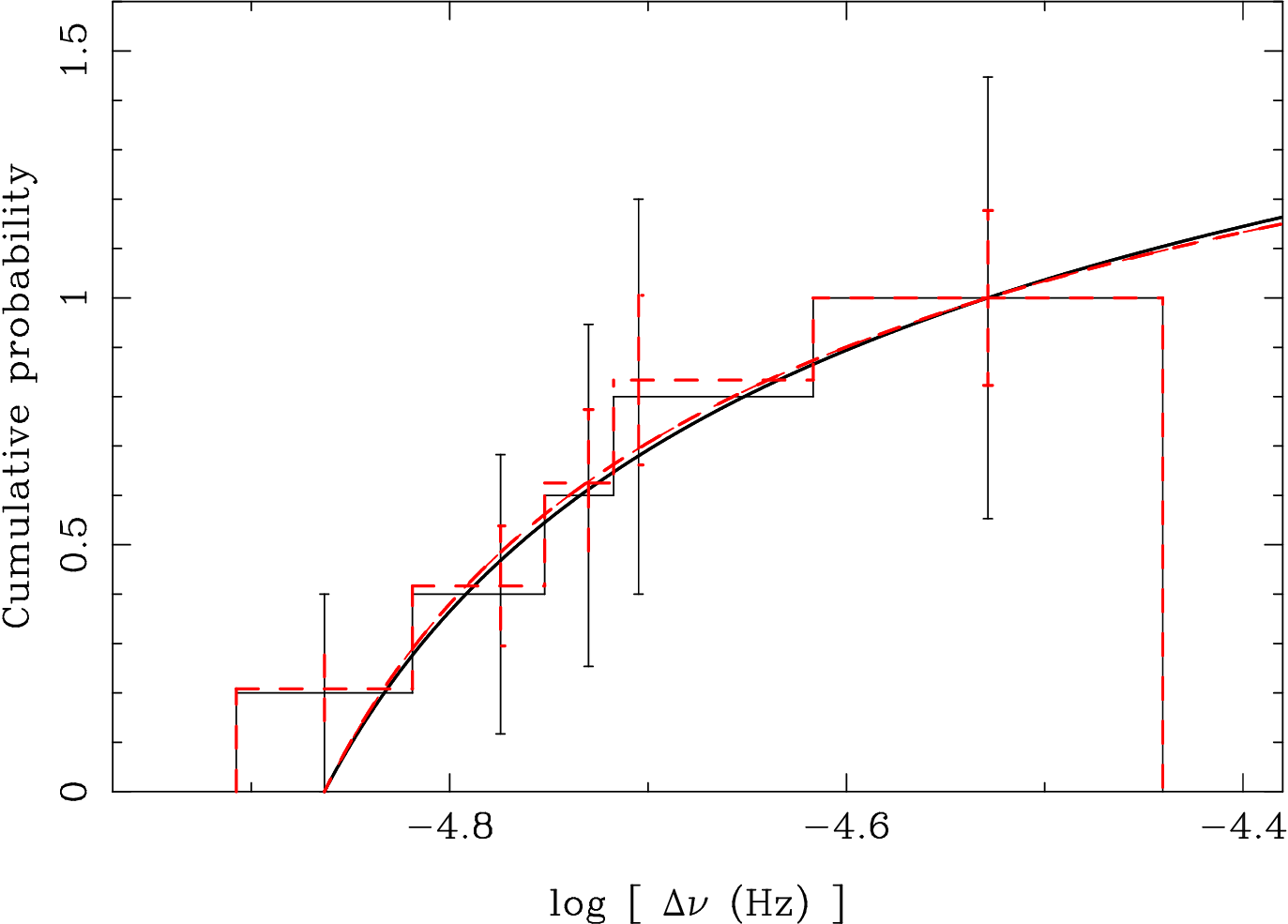} &
\includegraphics[angle=-90,width=7cm]{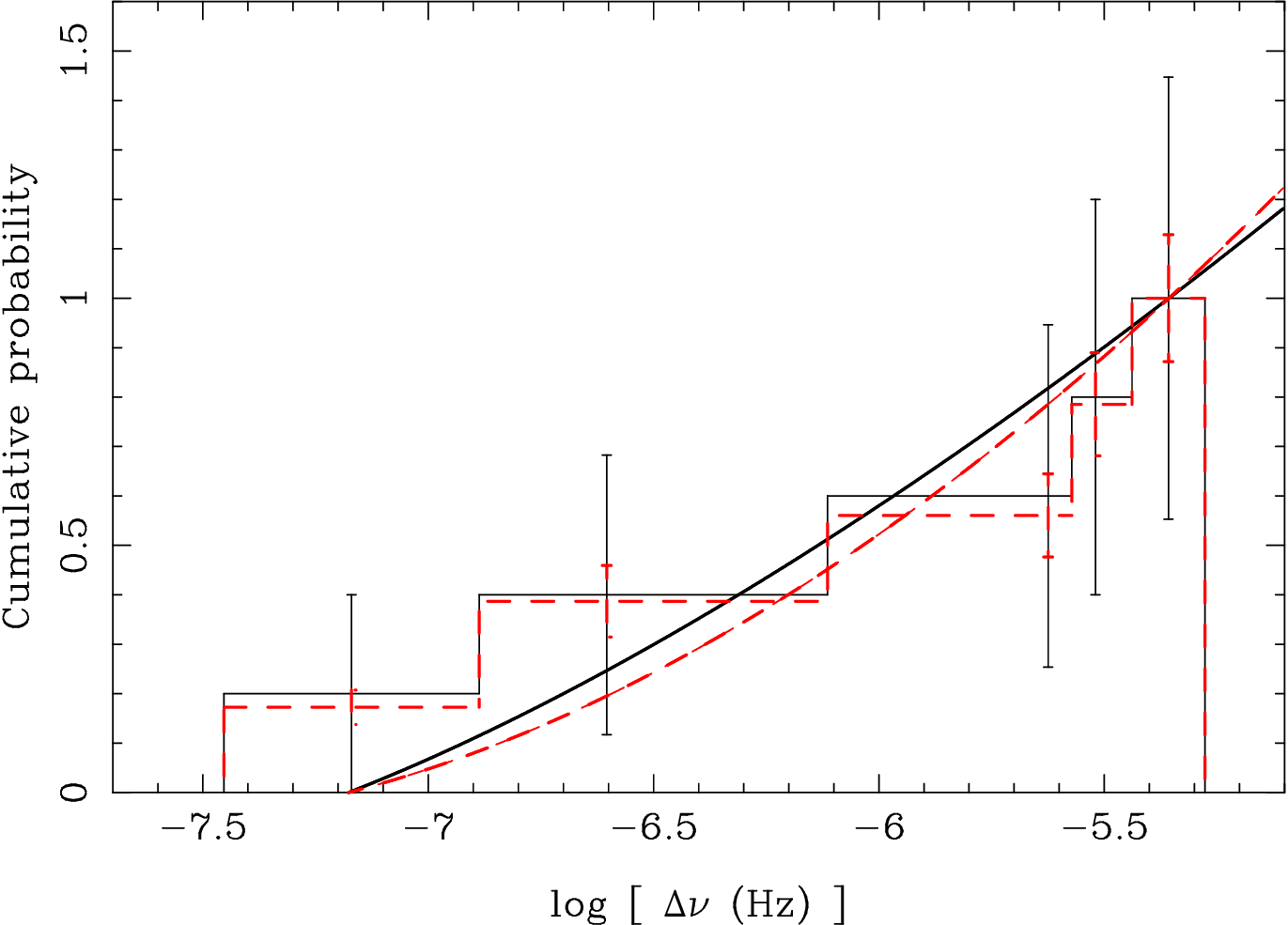}\\
\includegraphics[angle=-90,width=7cm]{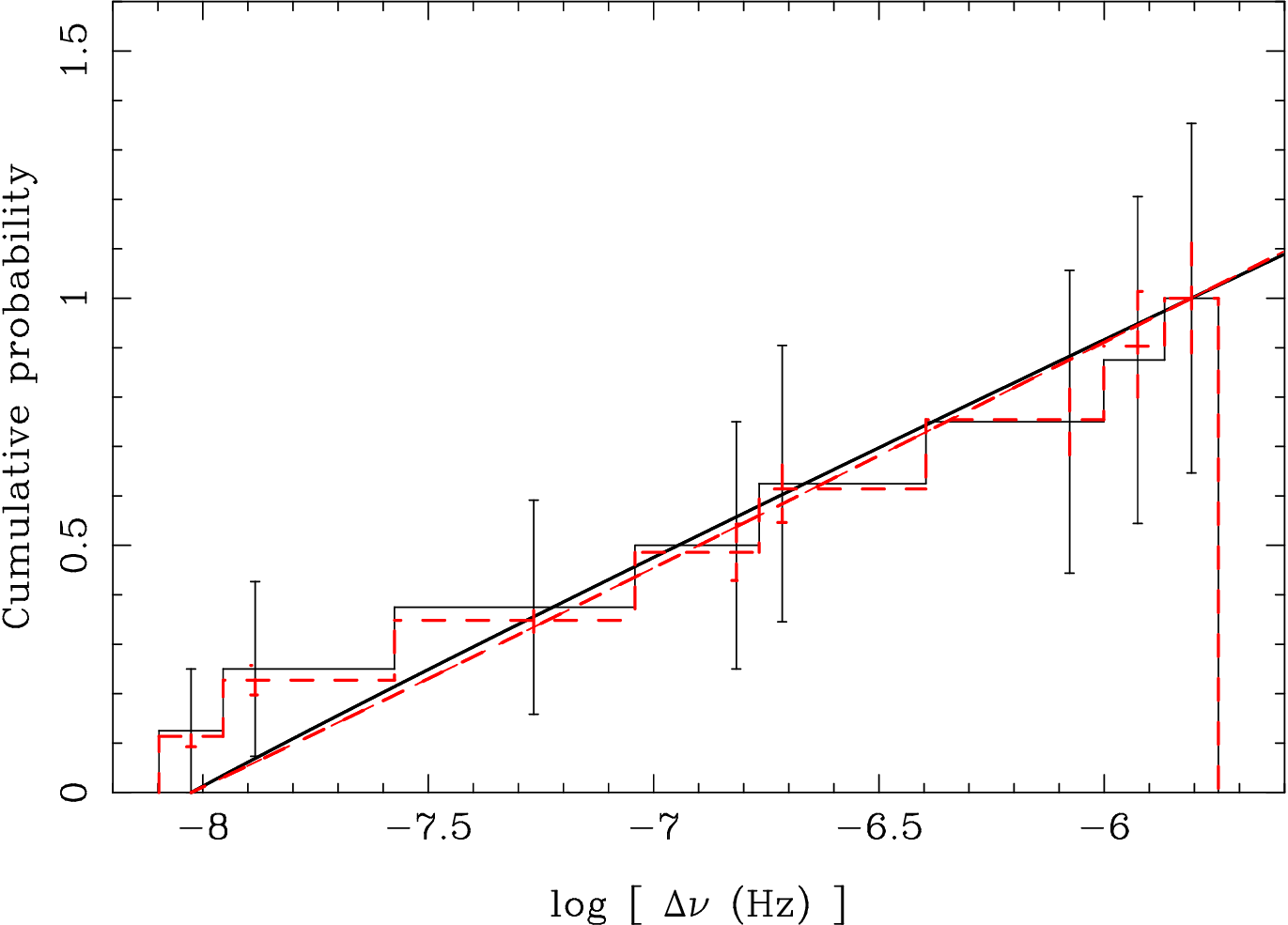} &
\includegraphics[angle=-90,width=7cm]{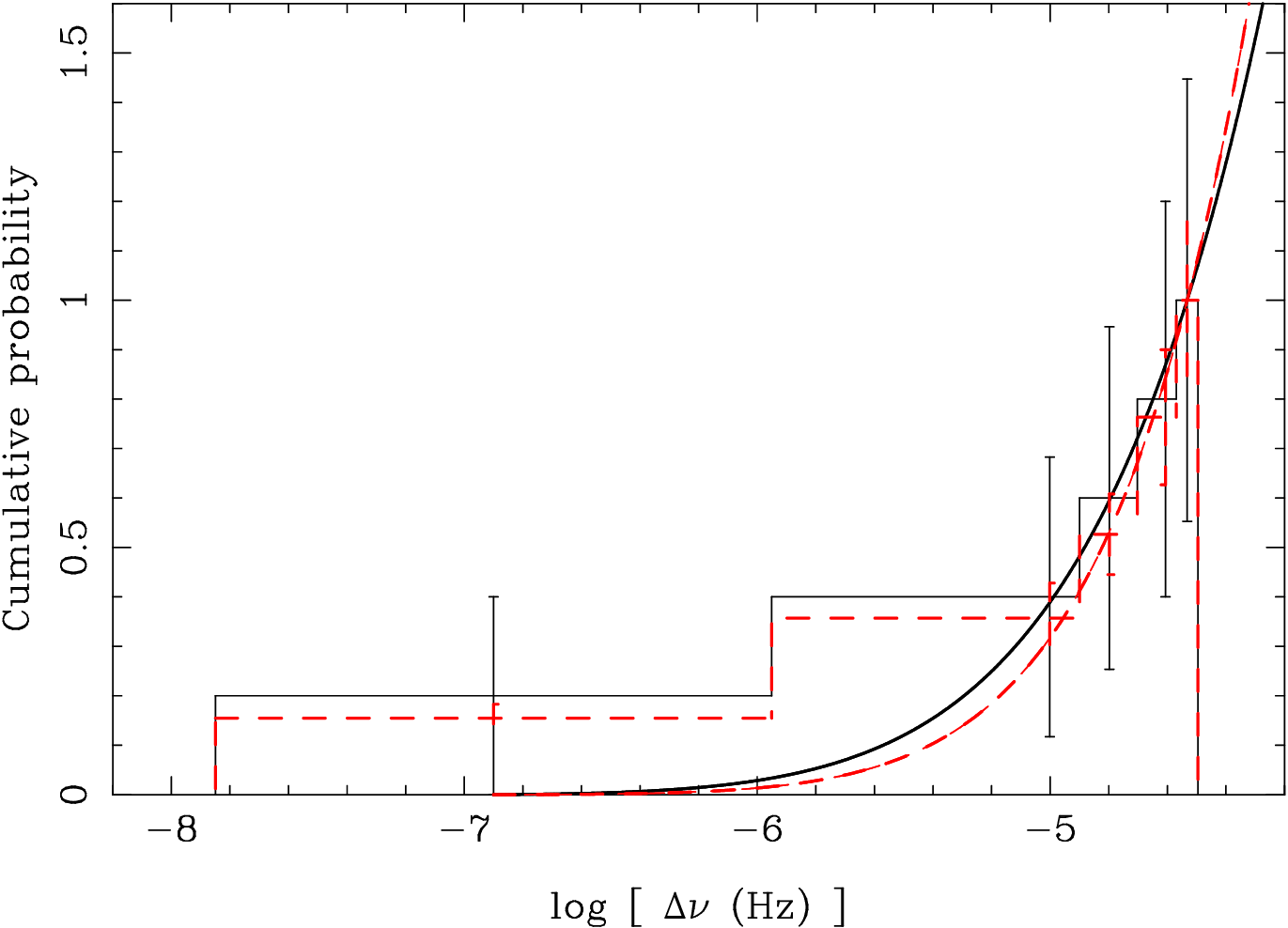}\\
\end{tabular}
\end{center}
\caption{The observed (black solid) and inferred (red dashed) glitch
  size cumulative distributions for PSRs~J1048$-$5832 and J1413$-$6141
  (first row), PSRs~J1420$-$6048 and J1740$-$3015 (second row),
  PSRs~J1801$-$2304 and J1801$-$2451 (third row). Curves are the
  fitted power-law models (see Table \ref{tab:ks} for
  parameters).}\label{fig:idistr_cdf}
\end{figure*}

After analysing PSR~J1341$-$6220 the same routine was implemented for
the other six pulsars each. Up until the final preparation of this
section, 294 realisations were obtained for PSR~J1048$-$5832, 209 for
PSR~J1413$-$6141, 201 for PSR~J1420$-$6048, 200 for PSR~J1740$-$3015,
200 for PSR~J1801$-$2304 and 195 for PSR~J1801$-$2451. In Figure
\ref{fig:idistr_p_tot} we present their glitch detection probability
densities. The lower detectabilities of small glitches ($\Delta\nu
\lesssim 10^{-7}$\,Hz) are commonly seen in the solutions of the noise
term. The small scale of the simulations results in large fluctuations
in the solutions of the epoch and multi-glitch terms and in the
solutions of the complete probability. In Figure \ref{fig:idistr_cdf}
CDFs of the observed glitch sizes for every pulsars are plotted as
solid bars, CDFs inferred with the detection probability densities and
observations are plotted as dashed bars. Table \ref{tab:ks} gives
results of the K-S tests. We see the corrections to the observed CDFs
with taking the detection probability densities into account are
insignificant for these cases. Therefore we can say the power
exponents measured are the values drawn from the data.

\begin{table*}
\caption{Parameters of the Kolmogorov-Smirnov tests for modelling the
  glitch size cumulative distributions of the seven pulsars with
  power-law functions. The columns give pulsar Jname, observed minimum
  and maximum glitch sizes, 68 per cent lower boundary, central value
  and 68 per cent upper boundary of the power exponent fitted for the
  observed distribution, 68 per cent false-alarm probability of the
  null hypothesis (the observed distribution and model are drawn from
  the same distribution), 68 per cent lower boundary, central value
  and 68 per cent upper boundary of the power exponent fitted for the
  inferred distribution, 68 per cent false-alarm probability of the
  null hypothesis (the inferred distribution and model are drawn from
  the same distribution).}\label{tab:ks}
\begin{center}
\begin{tabular}{cccrrrlcrrrl}
 & & & \multicolumn{4}{c}{Observed} & & \multicolumn{4}{c}{Inferred} \\
\cline{4-7} \cline{9-12} \\
PSR J & $\Delta\nu_{\rm min}$ & $\Delta\nu_{\rm max}$ & $s_-$ & $s$ & $s_+$ & $Q_{KS}$ & & $s_-$ & $s$ & $s_+$ & $Q_{KS}$ \\
 & (Hz) & (Hz) \\
\hline\\
J1048$-$5832 & $10^{-6.838}$ & $10^{-4.609}$ & $-$2.8 & 1.9 & 14.1 & 0.300 & & $-$2.3 & 2.0 & 12.9 & 0.219 \\
J1341$-$6220 & $10^{-7.211}$ & $10^{-5.073}$ & 0.0 & 0.7 & 2.1 & 0.147 & & $-$0.8 & $-$0.4 & 0.6 & 0.095 \\
J1413$-$6141 & $10^{-6.864}$ & $10^{-5.074}$ & $-$5.9 & $-$0.8 & 4.0 & 0.102 & & $-$8.3 & $-$1.8 & 5.7 & 0.302 \\
J1420$-$6048 & $10^{-4.528}$ & $10^{-4.863}$ & $-$3.8 & $-$1.3 & 0.8 & 0.001 & & $-$4.1 & $-$1.4 & 0.5 & 0.017 \\
J1740$-$3015 & $10^{-7.177}$ & $10^{-5.356}$ & $-$3.4 & 0.5 & 13.1 & 0.021 & & $-$3.1 & 0.9 & 13.7 & 0.029 \\
J1801$-$2304 & $10^{-8.026}$ & $10^{-5.806}$ & $-$1.7 & $-$0.1 & 2.6 & 0.002 & & $-$1.6 & 0.0 & 2.6 & 0.001 \\
J1801$-$2451 & $10^{-6.900}$ & $10^{-4.532}$ & $-$0.6 & 5.9 & 17.5 & 0.009 & & 0.0 & 7.4 & 24.6 & 0.001 \\ 
\hline\\ 
\end{tabular}
\end{center}
\end{table*}

\subsection{Significance for the avalanche model}\label{sec:disn:soc}
As described in Section \ref{sec:intro}, the avalanche model or the
general self-organised criticality expects power-law distributions for
glitch sizes. \citet{wm08} and \citet{mpw08} suggested the exponent is
a function of physical quantities e.g. stellar temperature and
strength of pinning forces etc. Therefore it becomes an essential
requirement for observers to determine the detection probability
densities for glitch events (such that power exponent embedded in the
data can be drawn).

However glitches have been found to violate scale-invariance. By
analysing the data set with high observing cadences for the Crab
pulsar, \citet{eas+14} found the smallest glitch detected was well
above the minimum detectable glitch size defined by the data set. They
concluded the Crab pulsar has a glitch size lower cut-off. 

\begin{figure}
\begin{center}
\begin{tabular}{c}
\includegraphics[angle=-90,width=8cm]{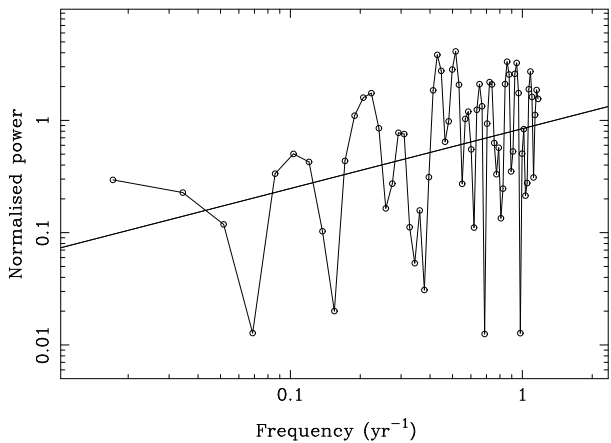}
\end{tabular}
\end{center}
\caption{The Lomb normalised periodogram for the glitch time series
  observed in PSR~J1341$-$6220. The straight line denotes the
  least-squares fit with $y=-0.07(16)+0.5(3)x$.}\label{fig:period}
\end{figure}

Glitch temporal behaviour is another point to compare with the
self-organised criticality. In Figure \ref{fig:period} we present the
Lomb normalised periodogram \citep{lom76,sca82} of the glitch time
series, the variation of glitch size as a function of glitch epoch, of
the seventeen glitches observed in PSR~J1341$-$6220. In making it, we
sampled the spectrum four times finer than at the conventional
interval, reciprocal of the time span, and derived the spectrum up
until two times the Nyquist frequency. We see, in logarithmic space,
the power demonstrates a tendency of increasing as frequency
increases. Although the derived detection probability densities
(Figure \ref{fig:idistr_J1341} upper panel) tell us we do not have the
glitch time series embedded in the data, the observed series do have
manifested the low-frequency (relative to the data span)
characteristic of the pulsar's glitch time series. A least-squares fit
to the periodogram showed the slope, so that the power exponent, is
$b=0.5(3)$. This is far from $-1$ the value of flicker noise. Our
analysis shows glitch temporal behaviour is not in agreement with the
self-organised criticality.

\section{Conclusion}\label{sec:conc}
In this work, we studied observational selection of neutron star
glitches in terms of detection probability using the fairly large
\citeauthor{ymh+13} data sets. The probability for detecting a glitch
event in pulsar timing observations is determined by the time of
occurrence of the glitch relative to entire observing span and
observing cadences, time of occurrence of preceding/subsequent
glitches relative to observing cadences and the strength of timing
noise. By implementing the Monte Carlo simulation with assuming
glitches distribute uniformly, we derived numerical solution of the
complete probability formula (equation \ref{eq:prob}) for the group
case that contained 157 pulsars. Using the obtained detection
probability densities and the observed distribution, we inferred the
aggregated glitch size probability distribution embedded in the
data. The inferred distribution is only a bit more massive than the
observed distribution for glitch sizes $\lesssim 4\times10^{-7}$\,Hz,
implying \citeauthor{ymh+13} have detected all detectable glitches in
the data using the manual method and we have well modelled the
detectabilities to glitches of the data sets and manual method. By
implementing Monte Carlo simulations in the same way, we derived
glitch detection probability densities for each of the seven pulsars
with glitch numbers $\geqslant 5$. With inferring the glitch size
distribution embedded in the data for the seven pulsars each, we
compared the power-law model of the inferred distribution with that of
the observed distribution. For PSRs~J1048$-$5832, J1413$-$6141,
J1420$-$6048, J1740$-$3015, J1801$-$2304 and J1801$-$2451, no
significant differences were seen. The most prominent difference
occurred for PSR~J1341$-$6220, the power exponent $s$ varied from the
observed $+0.7^{+1.4}_{-0.7}$ to the inferred $-0.4^{+1.0}_{-0.4}$. We
suggest observers determine the glitch detection probability. It helps
extract glitch distribution embedded in data and then plays a role in
studying theoretical models e.g. the avalanche model
\citep{wm08,mpw08}, the coherent noise model \citep{mw09} and the
Gross-Pitaevskii model \citep{wm11}. In addition, by deriving the Lomb
normalised periodogram for the glitch time series observed in
PSR~J1341$-$6220, we found the power exponent $b=0.5(3)$ is not what
flicker noise expects suggesting glitch phenomenon possesses different
temporal characteristic from the self-organised criticality.

However, the seventeen glitches observed in PSR~J1341$-$6220 are not
adequate to fully characterise the pulsar's glitch temporal
behaviour. More data are needed to measure the exponent $b$ more
accurately. Up till now, the power exponent $s$ has only been measured
for a few pulsars, intrinsic glitch size distribution has only been
inferred for the Crab pulsar \citep{eas+14}. Time scale of the rising
edge of pulse frequency at glitch has only been measured for a few
cases \citep[e.g.][]{lsp92,wbl01,dml02}, it has not been adequate to
study the distribution and correlation with glitch sizes. All of these
require more observations especially those with high sensitivity
instruments e.g. the Five-hundred-meter Aperture Spherical Telescope
(FAST) and Square Kilometre Array (SKA).
\section*{Acknowledgements}
This work is supported by the National Natural Science Foundation of
China (NSFC, No. 11403060), the Joint Research Fund in Astronomy
(U1531246) under cooperative agreement between the NSFC and Chinese
Academy of Sciences (CAS), the Strategic Priority Research Program
`The Emergence of Cosmological Structures' of the CAS
(No. XDB09000000), the International Partnership Program of the CAS
(No.114A11KYSB20160008) and the Strategic Priority Research Program of
the CAS (No. XDB23000000).

MY acknowledges Dr. R. N. Manchester for commenting the manuscripts;
Dr. G. Hobbs 1) for introducing software \textsc{ptasimulate} and
\textsc{temponest}, 2) together with Dr. J. B. Wang for introducing
the `pulse numbering' operation in \textsc{tempo2} and 3) for
commenting the initial manuscript; Dr. L. Lentati for
\textsc{temponest} set-up; Dr. K. J. Lee for helpful discussion. MY
especially acknowledges the anonymous referee for the comments which
greatly helped in improving the paper both in science and English
expression.

MY is extremely grateful to parents H. Q. He and G. C. Yu for their
encouragement.
%

%
\end{document}